\documentstyle[12pt]{article}

\newcommand{\om}{\omega}
\newcommand{\al}{\alpha}
\newcommand{\ep}{\epsilon}

\newcommand{\cO}{{\cal O}}

\newcommand{\deebar}{\bar{\partial}}

\newcommand{\df}{\stackrel{\rm def}{=}}

\newcommand{\msc}[1]{\mbox{\scriptsize #1}}
\newcommand{\dsp}{\displaystyle}

\newcommand{\bc}{\mbox{{\bf C}}}
\newcommand{\br}{\mbox{{\bf R}}}
\newcommand{\bz}{\mbox{{\bf Z}}}

\newcommand{\bsz}{\msc{{\bf Z}}}
\newcommand{\ba}{\mbox{{\bf a}}}
\newcommand{\bsa}{\msc{{\bf a}}}
\newcommand{\bm}{\mbox{{\bf m}}}
\newcommand{\bsm}{\msc{{\bf m}}}
\newcommand{\da}{\dot{a}}
\newcommand{\dal}{\dot{\alpha}}
\newcommand{\db}{\dot{b}}
\newcommand{\dbeta}{\dot{\beta}}
\newcommand{\cA}{{\cal A}}

\newcommand{\ket}[1]{{|#1\rangle}}

\newcommand{\sdprod}
  {\hbox{$\hspace{1mm}\rule{0.15mm}{2mm}\hspace{-0.7mm} \times$}}

\newcommand {\eqn}[1]{(\ref{#1})}

\newcommand{\mapru}[1]{\smash{\mathop{\hbox to 1cm{\rightarrowfill}}
\limits^{#1}}}
\newcommand{\maprd}[1]{\smash{\mathop{\hbox to 1cm{\rightarrowfill}}
\limits_{#1}}}
\newcommand{\maplu}[1]{\smash{\mathop{\hbox to 1cm{\leftarrowfill}}
 \limits^{#1}}}
\newcommand{\mapld}[1]{\smash{\mathop{\hbox to 1cm{\leftarrowfill}}
 \limits_{#1}}}
\newcommand{\maprud}[2]{\smash{\mathop{\hbox to 1cm{\rightarrowfill}}
 \limits^{#1}_{#2}}}
\newcommand{\maplud}[2]{\smash{\mathop{\hbox to 1cm{\leftarrowfill}}
 \limits^{#1}_{#2}}}

\newcommand{\lmapru}[1]{\smash{\mathop{\hbox to 1.5cm{\rightarrowfill}}
\limits^{#1}}}
\newcommand{\lmaprd}[1]{\smash{\mathop{\hbox to 1.5cm{\rightarrowfill}}
\limits_{#1}}}
\newcommand{\lmaplu}[1]{\smash{\mathop{\hbox to 1.5cm{\leftarrowfill}}
 \limits^{#1}}}
\newcommand{\lmapld}[1]{\smash{\mathop{\hbox to 1.5cm{\leftarrowfill}}
 \limits_{#1}}}
\newcommand{\lmaprud}[2]{\smash{\mathop{\hbox to 1.5cm{\rightarrowfill}}
 \limits^{#1}_{#2}}}
\newcommand{\lmaplud}[2]{\smash{\mathop{\hbox to 1.5cm{\leftarrowfill}}
 \limits^{#1}_{#2}}}

\newcommand{\cleqn}{\setcounter{equation}{0}}
\makeatletter
\@addtoreset{equation}{section}

\makeatother

\begin{document}
\vskip 7mm
%%% Title page %%%%%
\begin{titlepage}
 
 \renewcommand{\thefootnote}{\fnsymbol{footnote}}
 \font\csc=cmcsc10 scaled\magstep1
 {\baselineskip=14pt
 \rightline{
 \vbox{\hbox{hep-th/9903120}
       \hbox{UT-835}
       }}}

 \vfill
 \baselineskip=20pt
 \begin{center}
 \centerline{\Huge \rm $N=(0,4)$ Quiver $SCFT_2$ } 
 \vskip 2.5mm
 \centerline{\Huge  \rm and}
 \vskip 3mm
 \centerline{\Huge Supergravity on $AdS_3 \times S^2$}
 \vskip .8 truecm

 Yuji Sugawara\\
 {\sf sugawara@hep-th.phys.s.u-tokyo.ac.jp}

 \vskip .6 truecm
 {\baselineskip=15pt
 {\it Department of Physics,  Faculty of Science\\
  University of Tokyo\\
  Bunkyo-ku, Hongo 7-3-1, Tokyo 113-0033, Japan}
 }
 \vskip .4 truecm

 \end{center}

 \vfill
 \vskip 0.5 truecm

 \begin{abstract}
 \baselineskip 6.7mm

We study the proposed duality between the 5-dimensional
supergravity/superstring  on $AdS_3\times S^2$
and the 2-dimensional $N=(0,4)$ SCFT defined on the boundary
of $AdS$-space.
We construct explicitly the $N=(0,4)$ SCFT by imposing the ``quiver 
projection'' developed by Douglas-Moore \cite{DM}
on the $N=(4,4)$ SCFT of symmetric orbifold,
which is proposed  to be 
the dual of the 6-dimensional supergravity/superstring
on  $AdS_3\times S^3$. 

We explore in detail 
the spectrum of chiral primaries in this ``quiver $SCFT_2$''. 
We compare it with the Kaluza-Klein spectrum on $AdS_3\times S^2$
and check the consistency  between them.    
We further emphasize that orbifolding of bulk theory should {\em not\/}
correspond to orbifolding of the boundary CFT in the usual sense
of two dimensional CFT, rather corresponds to the quiver projection.
We observe that these are not actually equivalent with each other
when we focus on the  multi-particle states. 

\end{abstract}

 \setcounter{footnote}{0}
 \renewcommand{\thefootnote}{\arabic{footnote}}
\end{titlepage}

\newpage
\baselineskip 7mm

\section{Introduction}

\cleqn
\hspace*{4.5mm}

$AdS_3/CFT_2$-duality has been understood most successfully
among the several proposals 
of $AdS_{d+1}/CFT_d$-duality \cite{Mal,Witten1,GKP}
in  general dimensions. This is due to the fact that  the two dimensional 
conformal symmetry is very powerful to analyse the spectrum,
and three dimensional gravity with a negative cosmological term can
be quantized by recasting it to Chern-Simons theory \cite{Banados,Martinec}.
Among other things, one of the most remarkable success 
is the complete agreement between the BPS Kaluza-Klein (KK) spectrum 
of supergravity theory (SUGRA)
on $AdS_3\times S^3 \times M^4$  (with $M^4=T^4 ~\mbox{or}~ K3$) and 
the spectrum of chiral primary states of $N=(4,4)$ $SCFT_2$ with the
target manifold $Sym^N(M^4)$ \cite{deBoer1,deBoer2,MMS}.
However, analyses on the models with less supersymmetries
are still far from complete.
In this paper we shall focus on the correspondence between
the 5-dimensional SUGRA on $AdS_3\times S^2$ and the $N=(0,4)$ $SCFT_2$
defined on the boundary of $AdS_3$.

The most general examples which realize this geometry 
in the near horizon limit are given 
by considering the M-theory compactified 
on some Calabi-Yau 3-fold $M^6$ (including the cases $M^6=T^6, ~ 
T^2 \times K3$) \cite{Mal,MSW,Vafa}, 
and we can construct the boundary $N=(0,4)$ $SCFT_2$
by making M5-brane wind  around 
a suitable supersymmetric 4-cycle $P$ in $M^6$
with non-vanishing triple self-intersection: $P^3\neq 0$.
%(which corresponds to the ``4-charge model'' in the black-hole business.)
Although it is known that these $CFT_2$ can thoroughly reproduce 
the correct Bekenstein-Hawking entropy in the microscopic level,
it is  still a non-trivial problem  to analyse the detailed spectra 
of chiral primaries. The present work  is motivated by
this fact.

To resolve the difficulty to study the $N=(0,4)$ $SCFT_2$ derived 
from this  configuration of M5, we shall choose another way. That is,
we will work out in a T-dualized framework. 
We focus on the limited examples $M^6= T^6$ or $T^2 \times K3$
through this paper, and  can easily find that the above
brane configuration of M5 is equivalent by T-duality to the system
of D1-D5-TN5 in IIB string theory. (Here ``TN5'' means ``Taub-NUT
5-brane'', also sometimes called ``KK 5-brane'',  which is defined as 
a Taub-NUT solution in the transversal 4-dimensional space).
We take this brane configuration as the starting point to analyse 
the $N=(0,4)$ $SCFT_2$ following the several authors \cite{BBG,KLL,BL}. 
The advantage to do so is as follows:
The D1-D5 system leads to the well-known $N=(4,4)$ $SCFT_2$
we already mentioned,
which is finely understood at least as a CFT of symmetric orbifold,
%
%\footnote
% 
and putting these branes 
at the location of singularity of Taub-NUT ($\sim $ ALE) can be also 
described by the familiar procedure \cite{DM}, which is realized by
imposing suitable projection on the Chan-Paton (CP) indices. We shall call
this projection as the ``quiver projection'' through this paper.
In this manner  we can reduce our problem to a ``tractable'' one.
We will {\em define\/} the Hilbert space of 
the desired $(0,4)$ $SCFT_2$ by directly imposing the quiver projection on 
that of the $(4,4)$ $SCFT_2$ as a symmetric orbifold theory. 
Although the ``quiver (0,4) $SCFT_2$''
defined in this way  describes only the symmetric orbifold point 
of the moduli space, 
it is expected that we can analyse the spectrum which are {\em stable 
under marginal deformations\/} by means of this $SCFT_2$.
After carrying out the detailed analyses on chiral primaries in
the quiver $SCFT_2$, 
we will find out the complete agreement 
between the chiral primary states and the spectrum of KK excitations
in the 5d SUGRA on $AdS_3\times S^2$ under the Maldacena limit.

~

This paper will be organized as follows:

In section 2 we give a short review about some known results
of the  5-dimensional SUGRA with $AdS_3\times S^2$-geometry.
Particularly, we demonstrate the brane configuration on which our 
construction of the boundary $SCFT_2$ is based.

Our main discussions will be presented in section 3 and 4. 
We there  give our proposal of the boundary $N=(0,4)$ $SCFT_2$ 
as the one constructed by making use of the quiver techniques \cite{DM},
and explore the detailed spectrum of the chiral primary states. 
Our analysis of spectrum  consists of 
three parts; the single particles in 
the untwisted sector, the single particles
in the various twisted sectors, 
and lastly we discuss the spectrum of the general 
multi-particle states. We will find  out the complete agreement 
between the BPS KK spectrum of the $AdS_3\times S^2$-SUGRA and 
the chiral primaries in the quiver (0,4) $SCFT_2$. 
We will also emphasize, 
especially in relation to the analysis on multi-particles, 
that our  $SCFT_2$ defined by the quiver techniques is {\em not\/}
equivalent with that defined by the usual $\bz_q$-orbifoldization.
We observe that the results of SUGRA support the  
quiver (0,4) $SCFT_2$, not the one obtained by $\bz_q$-orbifoldization.

In section 5 we summarize our results and give a few comments about
open problems.

~

\section{5d SUGRA on $AdS_3\times S^2$}

\cleqn
\hspace*{4.5mm}

In this section we briefly survey some known results with respect to the 
5-dimensional SUGRA on $AdS_3\times S^2$ for 
later convenience. We begin with the simplest example:
M-theory compactified on $M^6= T^6$. We can also present 
the similar arguments for the case of $M^6= T^2 \times K3$.

For the 6-torus $M^6= T^6$, the M5-brane wrapped around a suitable SUSY
4-cycle which we previously mentioned means nothing but the three
intersecting 5-branes \cite{harmo}. Let $(x^5,x^6, \ldots , x^{11})$ be
the coordinates along the internal space $M^6$, and 
consider the following 5-brane configuration (we regard $x^0$
as the time coordinate);
\begin{itemize}
\item $q_1$ M5s extending along the (016789),
\item $q_2$ M5s extending along the (01567\,11),
\item $q_3$ M5s extending along the (01589\,11).
\end{itemize}  
The 11-dimensional metric corresponding to this brane configuration 
is given as follows \cite{harmo};
\begin{equation}
\begin{array}{l}
 ds^2_{11}= (H_1H_2H_3)^{2/3}\, \left\{
            (H_1H_2H_3)^{-1}(- (dx^0)^2 +(dx^1)^2) + (H_1H_2)^{-1} 
              ((dx^6)^2+ (dx^7)^2)    \right.          \\ 
   \left. ~~~~~     + (H_2H_3)^{-1} ((dx^5)^2+ (dx^{11})^2) + 
              (H_3H_1)^{-1} ((dx^8)^2+ (dx^9)^2) 
+  (dr^2+ r^2(d\theta^2 + \sin^2 \theta d\phi^2) ) \right\},
\end{array}
\end{equation}
where $(r,\theta, \phi)$ stands for 
the spherical coordinates along the $(x^2,x^3,x^4)$-directions which  
are transversal to all of the branes, and the harmonic functions $H_i$
in this transversal space should be defined as 
\begin{equation}
H_i(r)\df 1+ \frac{2\pi l_P q_i}{\om_2 r v^{1/3}}\equiv
1+ \frac{l_P q_i}{2 r v^{1/3}} .
\end{equation}
In the expressions of these harmonic functions 
$l_P $ means the 11-dimensional Planck length (which is defined, 
in our convention,
by $G_{11}= 16\pi^7 l_P^9$ from the 11-dimensional Newton constant $G_{11}$),
$\om_2 \equiv 4\pi$ means the area of unit 2-dimensional sphere
and we set the volume
of the internal space $M^6$ as $\mbox{Vol}(M^6)=v\,(2\pi l_P)^6$.
Under the Maldacena limit, that is,
\begin{equation}
D\equiv q_1q_2q_3 ~\longrightarrow~ \infty , 
~~~\mbox{with } l\equiv l_P \left(\frac{D}{v}\right)^{1/3} >> l_P  
~\mbox{ fixed,}
\label{mal limit}
\end{equation}
this metric reduces to that of $AdS_3\times S^2 \times M^6$;
\begin{equation}
ds^2= \left(\frac{l}{u}\right)^2\,du^2 + \left(\frac{u}{l}\right)^2 dx^+dx^-
+\frac{l^2}{4}\, (d\theta^2 + \sin^2 \theta d\phi^2)  + ds^2(M^6),
\end{equation}
where we set $u=\sqrt{2lr}$, $x^{\pm}= \pm x^0 + x^1$.

It is proposed \cite{Mal} that the boundary CFT for this example
should correspond to an $N=(0,4)$ $SCFT_2$ defined on the 
triple intersection of M5s ($x^0,x^1$-directions). This CFT 
is the simplest example among those studied in \cite{MSW,Vafa}
and was first considered in \cite{harmo} 
in the context of  the microscopic analysis on black-hole entropy. 
Its central charge can be evaluated under the ``large $N$ limit''
\eqn{mal limit} by counting the bosonic and fermionic degrees of
freedom. We can also evaluate it by means of the famous formula
by Brown-Hennaux \cite{BH} $\dsp c=\frac{3l}{2G_3}$ according to
the discussion given in \cite{Stro}.
Recalling that the 3-dimensional Newton constant $G_3$ is computed as 
\begin{equation}
\dsp \frac{1}{16\pi G_3}=\frac{1}{16\pi G_{11}}
\times \mbox{Vol}(M^6)\times  \mbox{Vol}(S^2)= 
\frac{1}{(2\pi)^8 l_P^9}\times (2\pi l_P)^6 v \times 
4\pi \left(\frac{l}{2}\right)^2 = \frac{v^{1/3}D^{2/3}}{4\pi l_P}, 
\end{equation}
we can immediately obtain the familiar  central charge $c=6D$ which 
reproduces the correct Bekenstein-Hawking entropy.

Although this evaluation seems satisfactory, this is not the whole
story in the $AdS_3/CFT_2$-correspondence. 
This SCFT defined on the triple intersection of 5-branes 
is not so powerful to analyse the detailed structure of spectrum.
Therefore, we shall take a different framework. It is easy to find 
the following equivalence by T-duality;
\begin{eqnarray}
M/(T^6(56789\,11)) \left\{\begin{array}{l}
        q_1~ M5 ~(016789)\\
        q_2~ M5 ~(01567 \,11) \\
        q_3~ M5 ~(015 89 \,11)
	     \end{array}\right.
&\cong & IIA/(T^5(56789)) \left\{\begin{array}{l}
		 q_1~  NS5 ~(016789)\\
                 q_2~ D4 ~(01567)\\
                 q_3~ D4 ~(01589)
			\end{array}\right. \nonumber \\
&\lmapru{T^5T^6T^7} &
IIB/ (\tilde{T}^5(56789)) \left\{\begin{array}{l}
		 q_1~ TN5 ~(016789)\\
                 q_2~ D1 ~(01) \\
                 q_3~ D5 ~(016789)
			\end{array}\right. ~~~. 
\label{T-dual}
\end{eqnarray}
In the last line  ``$TN5 ~(016789)$'' strictly   means
that the transversal space $(x^2,x^3,x^4,x^5)$ is the Taub-NUT 
space with the monopole charge $q_1$ (the $S^1$-fiber is along $x^5$). 
Lastly, let us consider the decompactification limit of Taub-NUT space, 
which means $(x^2,x^3,x^4,x^5)$ becomes  the ALE space 
$\dsp ALE(A_{q_1-1}) \sim \bc^2/\bz_{q_1}$.

In this way  we have led to the system of $Q_1\equiv q_1q_2$ $D1$ and 
$Q_1\equiv q_1q_3$ $D5$ wrapped around $T^4 (6789)$, which  are located at 
the singularity of the orbifold $\bc^2/\bz_{q_1}$ (the 2345th-directions). 
(The origin of the  extra factor  ``$q_1$'' 
in the brane charges is the existence of  mirror images by
the orbifold $\bz_{q_1}$-action.)
This brane configuration leads to the geometry $AdS_3 \times
S^3/\bz_{q_1}\times T^4$. Some analyses on this $AdS_3$ geometry 
were given  in \cite{BBG} and, the studies 
from the viewpoints of string theory
along the same line as \cite{GKS} were  
presented in \cite{KLL}. However, we shall take a route which 
is different from them and is rather analogous 
to the works \cite{KS,LNV,OT,Gukov}. 
The familiar techniques to put D-branes at the locations of orbifold 
singularities were invented by Douglas-Moore \cite{DM} and we shall
call this procedure the ``quiver projection'' in this paper.
In the next section we will construct the boundary SCFT by means of  this
projection from the well-known $(4,4)$-SCFT 
corresponding to the $Q_1$D1-$Q_5$D5 
system which has the near-horizon geometry $AdS_3 \times S^3 \times
M^4$. Furthermore, we will explore the spectrum of chiral primaries in this 
``quiver $(0,4)$-SCFT'', and will observe how we can get the complete 
agreement with the KK-spectrum of 5d SUGRA on $AdS_3 \times S^2$.
%%%
Here we should comment on the work \cite{BL}: it is there carefully
discussed that the perturbative degrees of freedom in such a quiver SCFT
as above are  completely reproduced also in the U-dualized framework 
of intersecting branes by incorporating  
the non-perturbative  BPS states of the  string junctions \cite{junction},
although the concrete analysis on the chiral primaries
as in the present paper are not given.
%%%

In the case of $M^6=T^2\times K3$ we can likewise T-dualize the orginal
system, at least for the elliptic K3\footnote
                 {Of course, the moduli space of elliptic K3 is merely a 
                 subspace of the full moduli space of K3. But our concern 
                  in this paper is to evaluate the spectrum
                 that does not depend on the generic value of  moduli.}.
We can again arrive at the $Q_1$D1-$Q_5$D5 system such that  $D5$s 
are wrapped around K3 and all of the branes are located at the orbifold 
point of $\bc^2/\bz_{q_1}$. 
We can also make up the boundary $(0,4)$ $SCFT_2$ by the quiver
projection, but there is one subtle point. 
It is known \cite{MSW,Vafa} that the central charge of this  boundary
CFT suffers  quantum correction, when $M^6$ is a curved manifold.
It seems difficult to realize this correction in the framework of our
quiver $SCFT_2$. However, since this correction becomes negligible 
if taking the Maldacena limit, 
we can expect it works in the K3 case, too.

To close this section let us present the BPS KK-spectra on 
$AdS_3 \times S^2 \times M^6$ SUGRA $(M^6=T^6,~ T^2\times K3)$ which are 
given in \cite{Larsen,deBoer1,KLL,FKM}.
These spectra should be encoded to the family of  the short 
representations of super Lie group
$SL(2,\br)_L \times SU(1,1|2)_R $ 
%(called ``Anti-de Sitter supergroup'' in some literatures) 
describing the symmetry of $AdS_3\times S^2$-SUGRA. 
The general irreducible representations of $SL(2,\br)_L \times SU(1,1|2)_R $
are  classified by three half integers 
$(h,\bar{h},\bar{j}) \in \frac{1}{2}\bz_{\geq 0}
\times \frac{1}{2}\bz_{\geq 0}\times \frac{1}{2}\bz_{\geq 0}$.
$h$ denotes the highest weight of $SL(2,\br)_L$-factor and 
$\bar{h},\bar{j}$ correspond to those of $SU(1,1|2)_R$.
The BPS condition can simply read as $\bar{h}=\bar{j}$.
We here express the each  short representation
by the notation $(\bar{j} \,;\, s)$, where $s\df h-\bar{h}\equiv
h-\bar{j}$ denotes the ``spin'', in the similar
manner as \cite{deBoer1}. Actually  $s$ is equal to the spin 
of the lowest component of  the BPS multiplet in the $AdS_3$-space. 

Under these preparations
the BPS KK-spectrum for $AdS_3\times S^2 \times T^6$ can be written
as follows;
\begin{equation}
\begin{array}{l}
 {\cal H}_{\msc{SUGRA}} (T^6) = {\cal H}^{(+)}_{\msc{SUGRA}}  (T^6)
       \oplus  {\cal H}^{(-)}_{\msc{SUGRA}} (T^6) ,   \\
~~~ {\cal H}^{(+)}_{\msc{SUGRA}} (T^6) = (1\,;\,2)+15(1\,;\,1)+14(1\,;\,0) \\
  \dsp ~~~~~ ~~~+\sum_{n=2}^{\infty}\, \left\{(n \,;\,2)
         +15(n\,;\,1) + 15(n\,;\,0)+ (n\,;\,-1)\right\}, \\
\dsp~~~ {\cal H}^{(-)}_{\msc{SUGRA}} (T^6) = 6(\frac{1}{2}\,;\,\frac{3}{2})
                               +14(\frac{1}{2}\,;\,\frac{1}{2}) \\
  \dsp ~~~~~ ~~~  +\sum_{n=2}^{\infty}\,
                \left\{6(n-\frac{1}{2}\,;\,\frac{3}{2}) 
         +20(n-\frac{1}{2}\,;\,\frac{1}{2})
       +6(n-\frac{1}{2}\,;\,-\frac{1}{2})\right\}.
\end{array} 
\label{KK T6} 
\end{equation}
In these expressions ${\cal H}^{(+)}_{\msc{SUGRA}}$ and  
${\cal H}^{(-)}_{\msc{SUGRA}}$ correspond to the spaces of 
the multiplets in which lowest component is a boson and a fermion
respectively.

For $AdS_3\times S^2 \times T^2 \times K3$, we can also write down the
spectrum in the similar manner;
\begin{equation}
\begin{array}{l}
 {\cal H}_{\msc{SUGRA}} (T^2\times K3) = 
{\cal H}^{(+)}_{\msc{SUGRA}} (T^2\times K3)
       \oplus  {\cal H}^{(-)}_{\msc{SUGRA}} (T^2\times K3)  ,  \\
~~~ {\cal H}^{(+)}_{\msc{SUGRA}} (T^2\times K3)
 = (1\,;\,2)+23(1\,;\,1)+22(1\,;\,0) \\
  \dsp ~~~~~ ~~~+\sum_{n=2}^{\infty}\, \left\{(n\,;\,2)
            +23(n\,;\,1) + 23(n\,;\,0)+ 
             (n\,;\,-1)\right\} , \\
~~~\dsp {\cal H}^{(-)}_{\msc{SUGRA}} (T^2\times K3)
         = 2(\frac{1}{2}\,;\,\frac{3}{2})
                               +42(\frac{1}{2}\,;\,\frac{1}{2}) \\
  \dsp ~~~~~ ~~~  +\sum_{n=2}^{\infty}\,
                \left\{2(n-\frac{1}{2}\,;\,\frac{3}{2}) 
         +44(n-\frac{1}{2}\,;\,\frac{1}{2})
       +2(n-\frac{1}{2}\,;\,-\frac{1}{2})\right\}.
\end{array}  
\label{KK K3}
\end{equation}

The mass spectrum of KK excitations  
is given from the values of ``conformal weights''
by the famous formula \cite{Witten1,MSW,deBoer1};
\begin{equation}
l^2m^2= (h+\bar{h})(h+\bar{h}-2) \equiv (2\bar{j}+s) (2\bar{j}+s-2).
\label{mass}
\end{equation}

~

%%%%%%%%%%%%%%%%%%%%%%%%%%%%%%%%%%%%%%%%%%%%%%%%%%%%%%%%%%%%%%%%%
\section{Quiver $N=(0,4)$ $SCFT_2$ as Boundary CFT}

\cleqn
\hspace*{4.5mm}

As was demonstrated in the previous seciton, the situation
we should study is summarized as follows;
\begin{itemize}
\item Q1 D1 extending along $(x^0,x^1)$.
\item Q5 D5 extending along $(x^0,x^1,x^6,x^7,x^8,x^9)$, and the space 
along $(x^6,x^7,x^8,x^9)$ is $M^4=T^4,\,K3$.
\item the space $ (x^2,x^3,x^4,x^5)$ is the orbifold $\bc^2/\bz_{q_1}$, and 
all of the branes sit at the orbifold singular point $x^2=\cdots = x^5=0$. 
\end{itemize}
Here we set $Q_1\equiv q_1q_2$, $Q_5\equiv q_1q_3$.

Notice that our orbifold group $\bz_{q_1}$ acts along the directions
transverse to all of the branes. It is different from the situations
intensively studied in \cite{DM}, in which the orbifold group
acts along $(x^6,x^7,x^8,x^9)$-directions 
and one  can gain  a natural interpretation 
of Kronheimer-Nakajima's theorem with respect to the instantons on 
ALE spaces \cite{KN} by means of the brane theory.

Nevertheless, we can apply the same techniques and can easily
write down the ``inner'' and ``outer'' quiver diagrams to encode the
matter contents. By this prescription we  can 
obtain  an $N=(0,4)$ supersymmetric theory
which is an example of the general models with $(0,4)$ SUSY
discussed in \cite{Witten2,Lambert}, called 
``ADHM gauged linear $\sigma$-model''.
In principle the IR limit of its Higgs branch
will provide us  the desired $N=(0,4)$ $SCFT_2$.

However, there are subtle points. In the quantum level it is not easy
to determine  exactly the  low energy effective theory of the ADHM 
linear $\sigma$-model, because we only have less supersymmetries.
Moreover, {\em even\/} in the cases with $N=(4,4)$ SUSY,
the famous proposal of  the target space $Sym^{Q_1Q_5}(M^4)$
is still a conjecture.
% except the easiest example $M^4=\br^4$.
No one succeeded in justifying this proposal completely from 
the viewpoint of two-dimensional quantum gauge theory.\footnote
           {In fact, there is a different proposal of the IR effective
            theory \cite{HW}. This has the same degrees of freedom,
             but a different discrete symmetry which should be
            gauged as orbifold $CFT_2$. So, it may lead to the different 
            spectrum of chiral primaries. } 

This $(4,4)$ $\sigma$-model on $Sym^{Q_1Q_5}(M^4)$ nonetheless
has a great success in $AdS_3/CFT_2$-duality 
\cite{MS,deBoer1,deBoer2,Dijkgraaf,MMS}.
So it seems to be the model worth taking as a good starting point.
In order to define our boundary $N=(0,4)$ $SCFT_2$, 
we shall directly carry out  the quiver projection on the 
$N=(4,4)$ orbifold $SCFT_2$ on $Sym^{Q_1Q_5}(M^4)$ {\em in place of 
taking the IR limit of the ADHM linear $\sigma$-model.}
This is our standpoint on which the analyses in this paper will be
based, and let us  first clarify the definition of Hilbert space  
of this theory.

~

\subsection{Hilbert Space of Quiver $SCFT_2$}

%\hspace*{4.5mm}

In this subsection we present 
the explicit definition of the quiver projection to
the Hilbert space of $Sym^{Q_1Q_5}(M^4)$ $SCFT_2$. 

For the time being we discuss the simpler example: $M^4=T^4$.
Let $\al, \dal = \pm $ be the spinor indices 
in the $ (x^2,x^3,x^4,x^5)$-space and 
$a,\da =\pm $ 
be those in the $ (x^6,x^7,x^8,x^9)$-space (compactified on $T^4$).

The description of the $N=(4,4)$  $SCFT_2$ 
on $Sym^{Q_1Q_5}(T^4)$ is elementary.
We have $4Q_1Q_5$ numbers of free bosons $X^{a\da}_{(AI)}$ and free fermions
$\psi^{\al\da}_{(AI)}$, $\bar{\psi}^{\dal\da}_{(AI)}$ 
($A=0,\ldots,Q_1-1$ ``color indices'', 
$I=0,\ldots,Q_5-1$ ``flavor indices'') as the fundamental
fields, which roughly correspond to the degrees of freedom of the
``1-5'' open strings. The superconformal symmetry is realized by the 
following currents (We only  write the left-mover.);
\begin{equation}
\begin{array}{l}
\dsp  T(z) = -\frac{1}{2}\sum_{AI}\, \ep_{ab}\, \ep_{\da\db} \, 
           \partial X_{(AI)}^{a\da} \, \partial X_{(AI)}^{b\db} 
          -\frac{1}{2}\sum_{AI} \,  \ep_{\al\beta}\, \ep_{\da\db}\,
             \psi_{(AI)}^{\al\da}\, \partial \psi_{(AI)}^{\beta\db}, \\
\dsp G^{\al a}(z) = i\sum_{AI}\, \ep_{\da\db}\, 
    \psi_{(AI)}^{\al\da} \, \partial X_{(AI)}^{a\db} , \\
\dsp J^{\al\beta}(z) = \frac{1}{2}\, \sum_{AI} \, \ep_{\da\db} \,
            \psi_{(AI)}^{\al\da} \, \psi_{(AI)}^{\beta\db}.
\end{array}
\label{SCA}
\end{equation}
These  generate the $N=4$ (small) superconformal algebra (SCA)
with central charge $c=6Q_1Q_5$. 
(The usual convention of $SU(2)$-current is $J^I(z)\equiv 
\sigma^I_{\al\beta} J^{\al\beta}(z)$ $(I=1,2,3)$.)
The subalgebra of ``zero-modes'' $\{ L_{\pm 1},\, L_0, \, 
G_{\pm 1/2}^{\al a}, \, J_0^{\al\beta} \}$ and their counterpart
of right-mover  composes the super Lie algebra
$SU(1,1|2)_L \times SU(1,1|2)_R$ corresponding to the symmetry supergroup
in the $AdS_3\times S^3$ geometry.  

According to the general theory of 
orbifold $CFT_2$ \cite{orbifold},
we have various twisted sectors labeled by the discrete symmetry group
$S_{Q_1Q_5}$. Each of them is defined as  the Fock space of bosonic and
fermionic  free fields with the twisted boundary conditions associated to 
$g\in S_{Q_1Q_5}$;
\begin{equation}
\Phi_{(AI)}(\sigma +2\pi ) = \Phi_{g((AI))}(\sigma), 
\label{twisted bc}
\end{equation}   
where $\Phi_{(AI)}(\sigma)$ denotes $X^{a\da}_{(AI)}$, 
$\psi^{\al\da}_{(AI)}$, $\bar{\psi}^{\dal \da}_{(AI)}$. 
The total Hilbert space of this orbifold theory should have the next
structure;
\begin{equation}
{\cal H}^{(4,4)} = \left(\bigoplus_{g\in S_{Q_1Q_5}}{\cal H}_{g} 
\right)^{S_{Q_1Q_5}}.
\label{(4,4) Hilbert space}
\end{equation}
Here ${\cal H}_{g}$ $(g\in S_{Q_1Q_5})$ denotes the Fock space of 
the $g$-twisted fields, 
and the superscript ``$S_{Q_1Q_5}$'' indicates the
$S_{Q_1Q_5}$-invariant subspace.
This Hilbert space \eqn{(4,4) Hilbert space} can be also expressed 
\cite{DMVV} as follows;
\begin{equation}
{\cal H}^{(4,4)} = \bigoplus_{[g]\in Y_{Q_1Q_5}} {\cal H}_{[g]}.
\label{(4,4) Hilbert space 2}
\end{equation}
Here the direct sum is taken over the set of all the  conjugate class
of $S_{Q_1Q_5}$ (which is identified with the set of Young tableaus of 
$Q_1Q_5$ boxes), and we set
\begin{equation}
{\cal H}_{[g]} = \bigoplus_{h \in [g]} ({\cal H}_{h})^{C(h)},
\label{(4,4) Hilbert space 3}
\end{equation}
where $C(h)\equiv \{g\in S_{Q_1Q_5}~:~ gh=hg\}$ 
denotes the centralizer of $h$.

Let us now consider the quiver projection of the $(4,4)$ $SCFT_2$.  
Since the Hilbert space of $(4,4)$ theory \eqn{(4,4) Hilbert space}
(or \eqn{(4,4) Hilbert space 2}) is defined as the Fock space of 
free fields with various boundary conditions \eqn{twisted bc},
it is enough to define the suitable projections \cite{DM} 
for each of the free fields
with these conditions.
% in order to construct
%the Hilbert space of $(0,4)$ theory.
To this aim
it may be  convenient to rewrite the indices $(AI)$ 
$(A=0,\ldots, Q_1-1,~ I=0,\ldots, Q_5-1)$ 
as $\left((a,j)\, (i,j')\right)$, where
$A=q_1a+j$ $(a=0,\ldots, q_2-1,~ j=0,\ldots,q_1-1)$, 
$I=q_1i+j'$ $(i=0,\ldots, q_3-1,~ j'=0,\ldots,q_1-1)$. 
Define the complex coordinates $Z^1=x^2+ix^3$, $Z^2=x^4+ix^5$
and set $\dsp {\cal Z}=\left(\begin{array}{cc}
    Z^1 & -\bar{Z}^2 \\
    Z^2 & \bar{Z}^1
\end{array}\right)$.
The actions of the isometry  group $SU(2)_L\times SU(2)_R$ along  
$(x^2,x^3,x^4,x^5)$-directions $(\sim S^3)$   are realized by
\begin{equation}
(g_L,g_R)\in SU(2)_L\times SU(2)_R ~:~ {\cal Z}~\longmapsto~
g_L{\cal Z}g_R.
\end{equation}
In particular, let us consider the $\bz_{q_1} \subset SU(2)_L$ action
generated by $\dsp g_L=\left(\begin{array}{cc}
		     \om & 0 \\
	              0& \om^{-1} \end{array}\right)$ 
$\dsp (\om\equiv e^{2\pi i/{q_1}})$, in other words,
\begin{equation}
Z^1 ~\longmapsto~\om Z^1 , ~~~ Z^2 ~\longmapsto~\om^{-1} Z^2.
\end{equation}
As was discussed in several papers (see for example \cite{Witten3}),
this $SU(2)_L\times SU(2)_R$ is no other than the R-symmetry group
of (4,4) $SCFT_2$ considered above. Therefore,
we can expect that  ``orbifolding''\footnote
               {The term ``orbifolding'' here is somewhat inaccurate, as 
                was remarked in \cite{LNV}. We will later discuss this point
                 carefully, especially relating it with 
                the analysis of multi-particle states. }
by this $\bz_{q_1}$-action breaks the SUSY of left-mover completely.

Under  these preparations the quiver projection associated to  
the orbifold group $\bz_{q_1}$ should be defined as
\begin{eqnarray}
 X^{a\da}_{((a,j+n),\,(i,j'+n))}&=& X^{a\da}_{((a,j),\,(i,j'))} ,
    \label{quiver X} \\
\psi^{\al \da}_{((a,j+n),\,(i,j'+n))}&=& \om^{n\al} \,
 \psi^{\al \da}_{((a,j),\,(i,j'))} ,  \label{quiver psi} \\
\bar{\psi}^{\dal \da}_{((a,j+n),\,(i,j'+n))}&=&
\bar{\psi}^{\dal \da}_{((a,j),\,(i,j'))} .  \label{quiver psi bar}
\end{eqnarray}
%where $\dsp \om\equiv e^{2\pi i/q_1} $.

It is not so difficult to check that this projection
actually breaks SUSY down  from (4,4)   to (0,4). 
We have no supercharges of left-mover  after this projection is performed. 
Obviously the bosonic and fermionic degrees of freedom
decrease to $1/q_1$ by them, and hence we have the central charge
$\dsp c=6\frac{Q_1Q_5}{q_1}\equiv 6q_1q_2q_3$ as expected. This
evaluation is also consistent with the Brown-Hennaux's formula
$\dsp c= \frac{3l}{2G_3}$, since $\dsp \frac{1}{G_3} \left( \propto
\mbox{Vol}(S^3)\right)$ becomes $1/q_1$ by 
the $\bz_{q_1}$-orbifoldization  of $S^3$.

Of course, all of the twisted boundary conditions \eqn{twisted bc}
are not necessarily compatible with the conditions of quiver projection
\eqn{quiver X}, \eqn{quiver psi}, \eqn{quiver psi bar}.
To clarify this point 
it may be useful to recast further the CP indices as follows; 
\begin{equation}
((a,j),\,(i,j')) \equiv (\tilde{j}, a, i \,; n) \equiv (\ba\,;n) , 
\end{equation}
where we set  $\tilde{j}\equiv j-j' \in \bz_{q_1}$ and $n\equiv j'$.
We also introduced the  abbreviated notation 
$\ba\equiv (\tilde{j}, a, i)\in \bz_{q_1}\times \bz_{q_2} \times \bz_{q_3}$.
In terms of this notation  we can rewrite the conditions of projections 
\eqn{quiver X}, \eqn{quiver psi}, \eqn{quiver psi bar} as
\begin{eqnarray}
 X^{a\da}_{(\bsa \,; n+m)}&=& X^{a\da}_{(\bsa \,;n)} ,
    \label{quiver X 2} \\
\psi^{\al \da}_{(\bsa \,; n+m)}&=& \om^{m\al} \,
 \psi^{\al \da}_{(\bsa \,;n)} ,  \label{quiver psi 2} \\
\bar{\psi}^{\dal \da}_{(\bsa \,; n+m)}&=&
\bar{\psi}^{\dal \da}_{(\bsa \,;n)} .  \label{quiver psi bar 2}
\end{eqnarray}
Let ${\cal S}$ be the subgroup of $S_{Q_1Q_5}$ composed of all the elements
of the form $(g,\bm)$ ($g\in S_{q_1q_2q_3}$, $\bm \equiv \{m_{\bsa}\} 
\in (\bz_{q_1})^{q_1q_2q_3} $), which naturally act on the CP indices 
in the next way;   
\begin{equation}
(g,\bm) ~: ~ (\ba \,; n) ~ \longmapsto~
(g(\ba) \,; n+m_{g(\bsa)}).
\label{local Z q1}
\end{equation}
Namely, we have 
${\cal S}\equiv S_{q_1q_2q_3}\sdprod (\bz_{q_1})^{q_1q_2q_3} $
(the symbol ``$\sdprod$'' standing for  the semi-direct product indicates
the relation $aba^{-1}b^{-1}\in  (\bz_{q_1})^{q_1q_2q_3}$ for
$\forall a\in S_{q_1q_2q_3}$, $\forall b\in (\bz_{q_1})^{q_1q_2q_3}$).
Notice that this $(\bz_{q_1})^{q_1q_2q_3}$-action is ``local'' in the
sense that it acts independently for each of  the $q_1q_2q_3$ numbers of 
CP indices $\ba \equiv (\tilde{j},a,i)$.
This fact will become important in our later discussions about
multi-particle states.

Under thess preparations one can easily find  that  
only the twisted sectors associated to
$g \in {\cal S}$ are compatible with the quiver projection.
(More precisely speaking, ${\cal S}$ is the maximal subgroup of $S_{Q_1Q_5}$
such that it has this property.)
By the same reason ${\cal S}$ is the discrete  symmetry group 
which we should ``gauge'' in the quiver projected theory.
Therefore we should define the Hilbert space for the quiver $SCFT_2$
as follows;
\begin{equation}
{\cal H}^{(0,4)} = \left(\bigoplus_{g\in {\cal S}}{\cal H}^{\msc{quiver}}_{g} 
\right)^{{\cal S}}, 
\label{(0,4) Hilbert space}
\end{equation}
where ${\cal H}^{\msc{quiver}}_{g}$ ($g\in {\cal S}$) is defined as
the Fock space of the $g$-twisted free fields 
with the quiver projections
\eqn{quiver X 2}, \eqn{quiver psi 2}, \eqn{quiver psi bar 2} imposed. 
Because the constraints
\eqn{quiver X 2}, \eqn{quiver psi bar 2}
imply the $(\bz_{q_1})^{q_1q_2q_3}$-part of the symmetry group 
${\cal S}$ trivially acts on the  right-mover, 
the Hilbert space of right-mover is the same as that 
of $Sym^{q_1q_2q_3}(M^4)$ $SCFT_2$. This fact seems
consistent with the observations in previous several works
\cite{harmo,Larsen}. However, the condition for the left-moving fermion
\eqn{quiver psi 2} means  the $(\bz_{q_1})^{q_1q_2q_3}$-action
on it  is non-trivial. 
It plays an quite important role in our later analysis of spectrum.  

Also in the case of  $M^4= K3$ we can likewise construct the Hilbert 
space of the quiver $SCFT_2$  at least as long as we regard  
$K3\cong T^4/\bz_2$, and thus,  we do not repeat the detailed arguments.

%%%%%%%%%%%%%%%%%%%%%%%%%%%%%%%%%%%%%%%%%%%%%%%%%%%%%%%%%%%%%%%%%%%%%%%%%%
~

\section{Chiral Primaries of Quiver (0,4) $SCFT_2$}

\cleqn
\hspace*{4.5mm}

In this section we consider the spectrum of chiral primaries in
our quiver (0,4) $SCFT_2$. In general the chiral primaries in the (0,4)
theory is obtained from the 1/4 BPS states in the (4,4) theory
(only the right-moving SUSY are preserved).
However, the analyses in the present paper 
are  performed only at the symmetric  orbifold point in the moduli space,
which is essentially a free CFT, and  
all the 1/4 BPS states at this point 
are not always stable under marginal deformations.

To clarify this point, remember that  
the analysis for the case $M^4= K3$ given in \cite{deBoer2}
counts only the 1/4 BPS states  {\em contributing  to the elliptic genus},
which is an index invariant under marginal deformations of CFT.
More recently  the similar analysis 
for $M^4=T^4$ based on the ``new SUSY index'' 
was given in \cite{MMS}. Their results imply that 
all the ``stable'' 1/4 BPS states contributing to these indices
with sufficiently small R-charges
(below the bound $\dsp \sim  \frac{Q_1Q_5}{4}$)
should be  either some descendants of (4,4)
chiral primaries (1/2 BPS) of the single particle type  or 
the multi-particle states, (later we clarify the precise meanings 
of the ``single particle'' and the ``multi-paritcle'' types)
and this aspect is consistent with the KK spectrum 
of SUGRA on $AdS_3\times S^3$.

In fact, there are many 1/4 BPS states which do not contribute
to the elliptic genus (in the case $M^4=T^4/\bz_2 \cong K3$) 
in our orbifold Hilbert space \eqn{(4,4) Hilbert space}.
For example, 
\begin{equation}
\begin{array}{l}
\dsp \sum_{(\bsa;n)}\, \psi_{(\bsa;n)\,-1/2}^{++}
\psi_{(\bsa;n)\,-1/2}^{-+} \ket{0}\otimes \bar{\ket{0}}, \\
\dsp \sum_{(\bsa;n)}\, \left\{ \psi_{(\bsa;n)\,-1/2}^{++}
\psi_{(\bsa;n)\,-1/2}^{--} 
- \psi_{(\bsa;n)\,-1/2}^{-+}
\psi_{(\bsa;n)\,-1/2}^{+-}   \right\} 
\ket{0}\otimes \bar{\ket{0}} , \\
\dsp \sum_{(\bsa;n)}\, \psi_{(\bsa;n)\,-1/2}^{+-}
\psi_{(\bsa;n)\,-1/2}^{--} \ket{0}\otimes \bar{\ket{0}}
\end{array}
\label{exceptions}
\end{equation}
are indeed 1/4 BPS states (and define chiral primaries in the (0,4)-theory,
since they survive after the quiver projection),
but they do not appear in the expansion  of elliptic genus\footnote
     {In fact, in \cite{deBoer1},  the structures of the 1/4 BPS states
     which have $h=1$, $j=0$, $\bar{h}\equiv \bar{j}=\mbox{arbitrary}$, 
     and can contribute to the elliptic genus
     are  completely examined. 
     The BPS states considered here \eqn{exceptions} 
     do not belong to this list.}.
Since we should count only the chiral primaries in (0,4) theory
which are stable under deformations of moduli, 
it is reasonable  to consider the quiver  projection  {\em only  
on the stable BPS states in the sense of the (4,4) theory\/}, 
that is, the BPS states contributing to the indices mentioned above.  

There may be a subtlety in this argument, since we do not 
directly consider  some index which is invariant
under marginal deformations {\em as (0,4) $SCFT_2$}.
To obtain  a rigid understanding of spectrum,
we will have to carry out a detailed  analysis based on such an index,  
and to this aim we  may perhaps
need to study  thoroughly the modular property of our (0,4) model.
It is beyond the scope of this paper, and  we would like to discuss 
this problem elsewhere. 

In any case we shall assume the validity of the above arguments and 
go on the  analysis of the chiral primaries.
Under this assumption we only have to look for the candidates 
of the stable (0,4) chiral primaries of the single particle type {\em among 
the $SU(1,1|2)_L$-descendants of (4,4) chiral primaries\/}, instead of 
searching for them in the total (0,4) Hilbert space \eqn{(0,4) Hilbert space}. 
This claim is based on the observations with respect to
the stable 1/4 BPS states in 
the (4,4) theory given in \cite{deBoer2,MMS} we mentioned above.
We believe the spectrum given  in the present paper is valid at least on
the generic points on the moduli space,
and it is enough for the purpose of checking  the correspondence 
between  the KK spectrum of the 5d SUGRA in bulk and 
the chiral primaries of the  boundary theory.

~

%%%%%%%%%%%%%%%%%%%%%%%%%%%%%%%%%%%%%%%%%%%%%%%%%%%%%%%%%%%%%%%%%%%%%%%%%%%%

\subsection{Spectrum of Single Particles : Untwisted Sector}

%\hspace*{4.5mm}

Let us first consider the untwisted sector 
in  the case $M^4=T^4$.
As was already explained, the (4,4) Hilbert space  in the untwisted sector 
is defined by imposing $S_{Q_1Q_5}$-invariance on the Fock space of 
the free bosons and fermions $\partial X_{(AI)}^{a\da}, ~ \deebar
X_{(AI)}^{a\da},~ \psi_{(AI)}^{\al\da},~  \bar{\psi}_{(AI)}^{\dal\da}$.

The chiral primary states are defined as follows;
\begin{equation}
\begin{array}{ll}
\dsp L_n\ket{\al} = \bar{L}_n \ket{\al} =0 & (\forall n \geq 1) ,\\
\dsp  G^{+ a}_r \ket{\al} = \bar{G}^{+ a}_r \ket{\al}=0  & 
 (\forall r \geq -\frac{1}{2}) , \\
\dsp  G^{- a}_r \ket{\al} = \bar{G}^{- a}_r \ket{\al}=0  & 
(\forall r \geq \frac{1}{2}) ,
\end{array}
\end{equation}
which corresponds to the short representations of 
$SU(1,1|2)_L \times SU(1,1|2)_R$.
By this definition we  inevitably have 
\begin{equation}
 L_0 \ket{\al}= J_0^3 \ket{\al}, ~  
\bar{L}_0 \ket{\al}= \bar{J}_0^3 \ket{\al} .
\end{equation} 
So, we can characterize the chiral primary states by the two quantum
numbers $(j,\bar{j})$ - spins of the two $SU(2)$-currents $J^{\al\beta}$, 
$\bar{J}^{\dal\dbeta}$.

As is well-known, the chiral primary fields, which corresponds to
the chiral primary states by the standard operator-state correspondence
in  unitary $CFT_2$, composes a ring by means of 
the simple operator products that has no singularity. This ring is
called ``chiral ring'', and from the viewpoints of $AdS_3/CFT_2$-duality
the generators of chiral ring are identified with the single particle
states \cite{MS} (see also \cite{deBoer1}). 
So, let us call them the ``chiral primaries of single
particle type'', and call the other operators in chiral ring
the ``multi-particle type''. 

By taking  the $S_{Q_1Q_5}$-invariance into account, 
we can explicitly construct all of the chiral primaries of 
single paritcle type in the untwisted sector;
\begin{equation}
\begin{array}{l}
\ket{\om^{(0,0)},(0)} = 
\ket{0}\otimes \bar{\ket{0}} ~~~\mbox{(NS-NS vacuum)} \\
\dsp \ket{\om^{(1,0)\da},(0)} = \sum_{A,I}\,\psi_{(AI)\,-1/2}^{+\da} 
\ket{0}\otimes \bar{\ket{0}} ,~~~
\dsp \ket{\om^{(0,1)\da},(0)} = \sum_{A,I}\,\bar{\psi}_{(AI)\,-1/2}^{+\da} 
\ket{0}\otimes \bar{\ket{0}} ,~~~ \\
\dsp \ket{\om^{(2,0)},(0)} 
= \sum_{A,I}\,\psi_{(AI)\,-1/2}^{++}\psi_{(AI)\,-1/2}^{+-} 
\ket{0}\otimes \bar{\ket{0}} , ~~~
\dsp \ket{\om^{(0,2)},(0)} = \sum_{A,I}\,\bar{\psi}_{AI\,-1/2}^{++}
\bar{\psi}_{AI\,-1/2}^{+-} 
\ket{0}\otimes \bar{\ket{0}} , \\
\dsp \ket{\om^{(1,1)\da\db},(0)} = \sum_{A,I}\,\psi_{(AI)\,-1/2}^{+\da}
 \bar{\psi}_{(AI)\,-1/2}^{+\db} \ket{0}\otimes \bar{\ket{0}}  , \\
\dsp \ket{\om^{(2,1)\da},(0)} = 
\sum_{A,I}\,\psi_{(AI)\,-1/2}^{++}\psi_{(AI)\,-1/2}^{+-} 
\bar{\psi}_{(AI)\,-1/2}^{+\da}\ket{0}\otimes \bar{\ket{0}} , \\
\dsp \ket{\om^{(1,2)\da},(0)} = \sum_{A,I}\,\psi_{(AI)\,-1/2}^{+\da}
\bar{\psi}_{(AI)\,-1/2}^{++}\bar{\psi}_{(AI)\,-1/2}^{+-} 
\ket{0}\otimes \bar{\ket{0}} , \\
\dsp \ket{\om^{(2,2)},(0)} 
= \sum_{A,I}\,\psi_{(AI)\,-1/2}^{++}\psi_{(AI)\,-1/2}^{+-}
\bar{\psi}_{(AI)\,-1/2}^{++}\bar{\psi}_{(AI)\,-1/2}^{+-} 
\ket{0}\otimes \bar{\ket{0}} .
\end{array}
\label{untwisted (4,4)}
\end{equation}
Needless to say, these correspond to the cohomology ring of $T^4$.

%%%%%%%%%%%%%%%%%%%%%%%%%%%%%%%%%%%%%%%%%%%%%%%%%%%%%%%%%%%%%%%%%%%%%%%%%%%
At this stage,
by perfoming  the quiver projection \eqn{quiver X}, \eqn{quiver psi},
\eqn{quiver psi bar} (or \eqn{quiver X 2}, \eqn{quiver psi 2},
\eqn{quiver psi bar 2}) and imposing the ${\cal S}$-invariance,  
we can immediately  obtain the (0,4) chiral primaries 
in the untwisted sector.
The most non-trivial part of this procedure  is 
the requirement of the $ (\bz_{q_1})^{q_1q_2q_3}$-invariance.  
This invariance has no meaning for the right-mover, 
since it is trivially satisfied because of 
\eqn{quiver X}, \eqn{quiver psi bar}. But, taking acount of  
the condition for $\psi^{\al \da}$ \eqn{quiver psi}, we find that 
only the states which are $SU(2)_L$-neutral
(modulo $\bz_{q_1}$) survives 
when the  $ (\bz_{q_1})^{q_1q_2q_3}$-invariance are imposed.
In this way we can obtain the following chiral primaries of 
the (0,4) theory;
\begin{equation}
\begin{array}{lll}
\ket{\om^{(0,0)},(0)}^{(0,4)}& =& 
\ket{0}\otimes \bar{\ket{0}} , \\
%%%
 \ket{\om^{(0,1)\da},(0)}^{(0,4)}& =& \dsp \sum_{(\bsa;n) }\,
\bar{\psi}_{(\bsa;n)\,-1/2}^{+\da} 
\ket{0}\otimes \bar{\ket{0}} \equiv q_1\sum_{\bsa}\,
\bar{\psi}_{(\bsa;0)\,-1/2}^{+\da} 
\ket{0}\otimes \bar{\ket{0}} , \\
%%%
 \ket{\om^{(2,0)},(0)}^{(0,4)} 
&=& \dsp \sum_{(\bsa;n)}\, \left\{ \psi_{(\bsa;n)\,-1/2}^{++}
\psi_{(\bsa;n)\,-1/2}^{--} 
+ \psi_{(\bsa;n)\,-1/2}^{-+}
\psi_{(\bsa;n)\,-1/2}^{+-}   \right\} 
\ket{0}\otimes \bar{\ket{0}} \\
%%%
 & \equiv&\dsp  q_1\sum_{\bsa}\,
\left\{ \psi_{(\bsa;0)\,-1/2}^{++}
\psi_{(\bsa;0)\,-1/2}^{--} 
+ \psi_{(\bsa;0)\,-1/2}^{-+}
\psi_{(\bsa;0)\,-1/2}^{+-}   \right\} 
\ket{0}\otimes \bar{\ket{0}} , \\
%%%
 \ket{\om^{(0,2)},(0)}^{(0,4)}& = &\dsp
\sum_{(\bsa;n)}\,\bar{\psi}_{(\bsa;n)\,-1/2}^{++}
\bar{\psi}_{(\bsa;n)\,-1/2}^{+-} 
\ket{0}\otimes \bar{\ket{0}}  \equiv 
q_1 \sum_{\bsa}\,\bar{\psi}_{(\bsa;0)\,-1/2}^{++}
\bar{\psi}_{(\bsa;0)\,-1/2}^{+-} 
\ket{0}\otimes \bar{\ket{0}} , \\
%%%
 \ket{\om^{(2,1)\da},(0)}^{(0,4)} &=& \dsp
\sum_{(\bsa;n)}\, \left\{ \psi_{(\bsa;n)\,-1/2}^{++}
\psi_{(\bsa;n)\,-1/2}^{--} 
+ \psi_{(\bsa;n)\,-1/2}^{-+}
\psi_{(\bsa;n) \,-1/2}^{+-}   \right\}  
\bar{\psi}_{(\bsa;n)\,-1/2}^{+\da}
\ket{0}\otimes \bar{\ket{0}} \\
 &\equiv& \dsp
q_1 \sum_{\bsa}\, \left\{ \psi_{(\bsa;0)\,-1/2}^{++}
\psi_{(\bsa;0)\,-1/2}^{--} 
+ \psi_{(\bsa;0)\,-1/2}^{-+}
\psi_{(\bsa;0) \,-1/2}^{+-}   \right\}  
\bar{\psi}_{(\bsa;0)\,-1/2}^{+\da}
\ket{0}\otimes \bar{\ket{0}} , \\
%%%
 \ket{\om^{(2,2)},(0)}^{(0,4)}& = &\dsp
\sum_{(\bsa;n)}\, \left\{ \psi_{(\bsa;n)\,-1/2}^{++}
\psi_{(\bsa;n)\,-1/2}^{--} 
+ \psi_{(\bsa;n)\,-1/2}^{-+}
\psi_{(\bsa;n)\,-1/2}^{+-}   \right\} \\
& & \dsp \hspace{2cm}\times \bar{\psi}_{(\bsa;n)\,-1/2}^{++}
\bar{\psi}_{(\bsa;n)\,-1/2}^{+-}
\ket{0}\otimes \bar{\ket{0}} \\
 & \equiv& \dsp
q_1 \sum_{\bsa}\, \left\{ \psi_{(\bsa;0)\,-1/2}^{++}
\psi_{(\bsa;0)\,-1/2}^{--} 
+ \psi_{(\bsa;0)\,-1/2}^{-+}
\psi_{(\bsa;0)\,-1/2}^{+-}   \right\} \\  
& & \dsp \hspace{2cm} \times
\bar{\psi}_{(\bsa;0)\,-1/2}^{++}
\bar{\psi}_{(\bsa;0)\,-1/2}^{+-}
\ket{0}\otimes \bar{\ket{0}} .
\end{array}
\label{untwisted (0,4)}
\end{equation}

%%%%%%%%%%%%%%%%%%%%%%%%%%%%%%%%%%%%%%%%%%%%%%%%%%%%%%%%%%%%%%%%%%%%%%%%%%%%%
The analysis for the case of $M^4=K3$ is also easy.
Since we now deal with only the BPS  states which are
stable under marginal deformations,
we may simply regard as $K3=T^4/\bz_2$. 
The constraint of $\bz_2$-invariance kills  
the odd cohomologies. Hence, in \eqn{untwisted (0,4)} only the states 
$\ket{\om^{(0,0)},(0)}^{(0,4)}$, $\ket{\om^{(2,0)},(0)}^{(0,4)}$, 
$\ket{\om^{(0,2)},(0)}^{(0,4)}$, $\ket{\om^{(2,2)},(0)}^{(0,4)}$ survive as 
the single particle chiral primaries in the untwisted sector. 
 
It should be remarked that in the orbifold  $T^4/\bz_2$,
we further have the $\bz_2$-twisted sectors  which 
correspond to the blow-up modes of 16 fixed points. 
In fact we must take account of them to recover the full cohomology of $K3$.
However, because all the contributions from these twisted sectors are 
$(1,1)$-cohomologies, we get no chiral primaries from
these sectors (the $SU(2)_L$-modules for them do not include neutral states). 
Of course, this is not the case when we deal with the twisted sectors 
of symmetric orbifold at the same time. We will later observe how 
the chiral primaries corresponding to these extra $(1,1)$-forms appear 
in the twisted sectors of symmetric orbifold.

~

%%%%%%%%%%%%%%%%%%%%%%%%%%%%%%%%%%%%%%%%%%%%%%%%%%%%%%%%%%%%%%%%%%%%%
\subsection{Spectrum of Single Particles : Twisted Sectors}

\hspace*{4.5mm}

The analysis for the twisted sectors is more complicated than that for
the untwisted sector.
As we observed before, we can suppose that 
each twisted sector of the symmetric orbifold corresponds to
a Young tableau $(n_1,\ldots,n_l)$, 
and this is decomposed into a set of $\bz_{n_i}$-twisted sectors,
each of which can  describe  a single-particle state 
with a large value of R-charge (see for example \cite{MS}).

Since we are here concerned about single particle states,  
we only have to learn how to describe the  $\bz_{p}$-twisted
sector which corresponds to the blow-up modes
of one of the orbifold singularities  such 
that $X^{a\da}_{(A_0I_0)}=\cdots = X^{a\da}_{(A_{p-1}I_{p-1})}$.
It is known that this can be elegantly but somewhat abstractedly
described  on some Fock space \cite{VW}. 
Namely, one can  prepare the bosonic and 
fermionic oscillators (in Ramond sector) 
associated to all the even and odd cohomology elements 
of $M^4$ respectively.
The oscillators of the type 
$\al_{-1}(\om)$ $(\forall \om \in H^*(M^4))$ correspond to 
the untwisted sector and those of the type $\al_{-p}(\om)$ 
describe the $\bz_p$-twisted sector.
Although this description has many excellences,
it may be more useful for our purpose
to describe ``concretely'' the chiral primaries
in terms of the $\sigma$-model variables themselves.

We must introduce the 
 ``$\bz_{p}$-twisted string'' or  the ``long stirng with 
length $p$'' to describe this sector\footnote
     {In this paper we are using the term ``long string'' 
      in the sense which  is different from  that given in 
      \cite{KS,HS}.}. We label the objects in this sector by
the index ${\cal A}\equiv  \left[(A_0I_0),\ldots ,(A_{p-1}I_{p-1}) \right]$, 
$A_i=0,\ldots,Q_1-1,~ I_i=0,\ldots,Q_5-1$, such that 
$(A_iI_i)\neq(A_jI_j)$ for  $\forall i\neq j$.
The long string coordinates for this sector (defined on the world-sheet
$0\leq \sigma \leq 2\pi $ which is rescaled from the
$p$-times one $0\leq \sigma \leq 2\pi p $)
is given as follows
($\Phi$ means each field $X^{a\da}$, $\psi^{\al\da}$, 
$\bar{\psi}^{\dal\da}$);
%\footnote
%   {Here we renormalize the ``length'' of the long  string 
%      from the $p$-times one 
%     $0\leq \sigma \leq 2\pi p$
%     to the usual one $0\leq \sigma \leq 2\pi$. 
%      So, our string coordinates have
%     the integral mode expansions as usual.
%     If we measure the energies of modes by $\hat{L}_0$ defined below, 
%    we have the fractional mode expansions.};
\begin{equation}
\Phi_{{\cal A}}(\sigma) = \Phi_{(A_r I_r)}(p\sigma - 2\pi r),
~~~\left(\frac{2\pi r}{p} \leq \sigma \leq \frac{2\pi (r+1)}{p}\right).
\label{long string}
\end{equation} 
These variables 
$X^{a\da}_{{\cal A}}$, $\psi^{\al\da}_{{\cal A}}$, 
$\bar{\psi}^{\dal\da}_{{\cal A}}$ can produce
the superconformal currents of the central charge $c=6$ 
in the same way as the untwisted sector \eqn{SCA}.
However, the condition of $\bz_{p}$-invariance which should be imposed
on the physical Hilbert space leaves us only the modes $n\in p\bz$ for
the bosons and the Ramond fermions, and the modes 
$\dsp n\in p\left(\frac{1}{2}+\bz \right)$ for the NS femions.
By this moding out, we can obtain the superconformal
currents $\{\hat{L}_n,\hat{G}_r^{\al a}, \hat{J}^{\al\beta}\}$ 
properly describing this $\bz_p$-twisted sector
\cite{FKN,BHS} (Here we only write the NS sector.);
\begin{equation}
\begin{array}{l}
\dsp \hat{L}_n = \frac{1}{p} L_{np} + \frac{p^2-1}{4p}\delta_{n0},   \\
\dsp \hat{G}_r^{\al a} =\left\{\begin{array}{ll}
               \frac{1}{\sqrt{p}} G_{pr}^{\al a \,(NS)} &  (p=2k+1) \\
	       \frac{1}{\sqrt{p}} G_{pr}^{\al a \,(R)} &  (p=2k)
  	\end{array} \right. , \\
\dsp \hat{J}_n = J_{np} .
\end{array}
\label{hat}
\end{equation}
These generate  the $N=4$ SCA with $c=6p$.
The anomaly term in the expression of $\hat{L}_n$ 
essentially corresponds to the 
Schwarzian derivative of the conformal 
mapping $z~\longmapsto ~ z^p$).

Observing these definitions \eqn{hat}, we can find that 
the NS vacuum $\ket{0,\cA}$ 
of $\bz_p$-twisted sector (corresponding to the twist
field which creates a cut on the world-sheet 
by the state-operator correspondence)
should possess the following properties;
\begin{itemize}
 \item $p=2k+1$
\begin{equation}
\begin{array}{l}
\dsp \hat{L}_0\ket{0,{\cal A}}= \frac{p^2-1}{4p}
\ket{0,{\cal A}},  \\
\dsp \hat{J}^3_0\ket{0,{\cal A}}= 0 .
\end{array}
\end{equation}
 \item $p=2k$
\begin{equation}
\begin{array}{l}
\dsp \hat{L}_0\ket{0,{\cal A}}= \left(\frac{p^2-1}{4p}
+\frac{1}{4p} \right)\ket{0,{\cal A}} 
\equiv \frac{p}{4}\ket{0,{\cal A}},  \\
\dsp \hat{J}^3_0\ket{0,{\cal A}}= -\frac{1}{2} \ket{0,{\cal A}} .
\end{array}
\end{equation}
\end{itemize}  
As one can find in \eqn{hat}, when $p$ is even, the NS supercurrent 
$\hat{G}_r^{\al a}$ is made up of the Ramond one before 
imposing the $\bz_p$-invariance. 
The extra vacuum energy ``$\dsp \frac{1}{4p}$''
and the extra R-charge  ``$\dsp -\frac{1}{2}$'' 
in the case of $p=2k$ originate from this fact.

The chiral primaries for this $\bz_p$-twisted sector are defined 
by the conditions; 
\begin{equation}
\begin{array}{ll}
\dsp \hat{L}_n \ket{\al}=\bar{\hat{L}}_n \ket{\al}=0 &
(\forall n \geq 1) , \\
\dsp  \hat{G}^{+ a}_r \ket{\al} = \bar{\hat{G}}^{+ a}_r \ket{\al}=0 & 
 (\forall r \geq -\frac{1}{2}) , \\
\dsp  \hat{G}^{- a}_r \ket{\al} = \bar{\hat{G}}^{- a}_r \ket{\al}=0 , & 
(\forall r \geq \frac{1}{2}) .
\end{array}
\label{hat CP}
\end{equation}
It may be slightly surprizing that the NS vacuum $\ket{0,{\cal A}}$
itself is {\em not\/} a chiral primary according to this conditions
\eqn{hat CP}.
By recalling the definitions of supercurrents $\hat{G}^{\al a}_r$ \eqn{hat},
we can find out that we must ``fill'' the NS vacuum 
with   all the fermionic oscillators
$\psi^{+\da}_{{\cal A}\, r}$  with $\dsp r< \frac{p}{2}$ to 
make it satisfy the conditions \eqn{hat CP}. 
Namely, all the possible chiral primaries 
in the $\bz_p$-twisted sector are written in the following form;  
\begin{equation}
\ket{\om^{(q,\bar{q})},(p-1)}=\sum_{{\cal A}}\,  \ket{\om^q,{\cal A}}\otimes
\overline{\ket{\om^{\bar{q}},{\cal A}}},
\end{equation}
where the summation should be taken over all the possible 
$\bz_p$-twisted strings labeled by $\cal A$, 
and  the left(right)-moving parts are defined by
\begin{itemize}
\item $p=2k+1$
\begin{equation}
\begin{array}{l}
\dsp  \ket{\om^0,{\cal A}}= \prod_{i=0}^{k-1}\,
\psi^{++}_{{\cal A}\, -\left(\frac{1}{2}+i \right)}
\psi^{+-}_{{\cal A}\, -\left(\frac{1}{2}+i \right)} \ket{0,{\cal A}}, \\
\dsp  \ket{\om^{1\da},{\cal A}}= \psi^{+\da}_{{\cal A}\, -\frac{p}{2}}
\ket{\om^0,{\cal A}} , \\
\dsp  \ket{\om^{2},{\cal A}}= \psi^{++}_{{\cal A}\, -\frac{p}{2}}
\psi^{+-}_{{\cal A}\, -\frac{p}{2}}
\ket{\om^0,{\cal A}} , \\
\end{array}
\end{equation}
%%%
\item $p=2k$
\begin{equation}
\begin{array}{l}
\dsp  \ket{\om^0,{\cal A}}= \prod_{i=0}^{k-1}\,
\psi^{++}_{{\cal A}\, -i}
\psi^{+-}_{{\cal A}\, -i} \ket{0,{\cal A}}, \\
\dsp  \ket{\om^{1\da},{\cal A}}= \psi^{+\da}_{{\cal A}\, -\frac{p}{2}}
\ket{\om^0,{\cal A}} , \\
\dsp  \ket{\om^{2},{\cal A}}= \psi^{++}_{{\cal A}\, -\frac{p}{2}}
\psi^{+-}_{{\cal A}\, -\frac{p}{2}}
\ket{\om^0,{\cal A}} . \\
\end{array}
\end{equation}
\end{itemize}
It is easy to check that\footnote
   {This means that the chiral primary state $\ket{\om^{(q,\bar{q})},(p-1)}$
   is identified with the element ``$\al_{-p}(\om^{(q,\bar{q})})\ket{0}$'' 
  of the fock space translating 
  the cohomologies of the Hilbert scheme \cite{VW,MS}.}
\begin{equation}
\begin{array}{l}
\dsp  \hat{L}_0 \ket{\om^{(q,\bar{q})},(p-1)} = \hat{J}^3_0 
\ket{\om^{(q,\bar{q})},(p-1)} = 
\frac{q+p-1}{2} \ket{\om^{(q,\bar{q})},(p-1)} , \\
\dsp  \bar{\hat{L}}_0 \ket{\om^{(q,\bar{q})},(p-1)} = \bar{\hat{J}}^3_0 
\ket{\om^{(q,\bar{q})},(p-1)} =
 \frac{\bar{q}+p-1}{2} \ket{\om^{(q,\bar{q})},(p-1)}.
\end{array}
\end{equation}

~

%%%%%%%%%%%%%%%%%%%%%%%%%%%%%%%%%%%%%%%%%%%%%%%%%%%%%%%%%%%%%%%%%%%%%%%%%
Now, let us consider the reduction to the (0,4) theory.
To observe how the quiver projection works on the twisted sectors, 
it may be useful to recast 
the long string variables $\Phi_{\cA}$ defined by \eqn{long string} as
follows (Again $\Phi_{\cA}$ means $X_{\cA}^{a\da}$, 
$\psi_{\cA}^{\al\da}$, $\bar{\psi}^{\dal\da}_{\cA}$);
\begin{equation}
\Phi_{{\cA}}^{(s)}(\sigma) \df \sum_{m\in \bsz}\, \Phi_{\cA \, pm+s}
e^{i(m+\frac{s}{p})\sigma} ~~~(s=0,\ldots,p-1), 
\end{equation}
where we assume the mode expansions of $\Phi_{\cA}$ is
\begin{equation}
\Phi_{{\cA}}(\sigma)= \sum_{m\in \bsz}\, \Phi_{\cA \, m} e^{im\sigma} .
\end{equation}
(As we already observed, when $p=2k+1$, 
 the fermionic fields $\psi_{\cA}^{\al\da}$,
$\bar{\psi}^{\dal\da}_{\cA}$
should have the half-integral mode expansions. It is straightforward 
to generalize our discussions here to this case, and we omit it.)
It is obvious that the set of  ``twisted fields'' 
$\{\Phi_{\cA}^{(s)}\}_{s=0,\ldots,p-1}$ describes the same degrees 
of freedom as $\Phi_{\cA}$ does.
These should have the monodromies;
\begin{equation}
\Phi_{\cA}^{(s)}(\sigma+2\pi)= \om_p^s \Phi_{\cA}^{(s)}(\sigma)~~~
(\om_p\equiv e^{\frac{2\pi i}{p}}) ,
\end{equation}
and this  implies the following formula of the Fourier transformation;
\begin{equation}
\left\{
\begin{array}{l}
\dsp  \Phi_{\cA}^{(s)}(\sigma) = \sum_{r=0}^{p-1}
\Phi_{(A_rI_r)}(\sigma)\om_p^{-rs} \\
\dsp \Phi_{(A_rI_r)}(\sigma) = \frac{1}{p}\sum_{s=0}^{p-1}
 \Phi_{\cA}^{(s)}(\sigma)\om_p^{rs} 
\end{array}
\right. .
\end{equation}
 
%The meaning of the quiver projections for $X^{a\da}$ \eqn{quiver X}
%and $\bar{\psi}^{\dal\da}$ \eqn{quiver psi} 
%is very easy. As was already mentioned,
%these simply reduce the degrees of freedom from 
%$Sym^{Q_1Q_5}(M^4)$ to  $Sym^{q_1q_2q_3}(M^4)$. 

As we showed in the previous section, in the quiver projected theory
only the $g\in {\cal S}$ twisted sectors are possible. 
Furthermore, according to our assumption, we only have to 
consider the projection on  the $SU(1,1|2)_L$-descendants of (4,4)
chiral primaries. 
These lead to  the fact that 
we only have to deal with the twisted sectors of the type;
${\cal A}\equiv 
 \left[(A_0I_0),\ldots ,(A_{p-1}I_{p-1}) \right] \equiv
 \left[(\ba_0;n_0),\ldots , 
(\ba_{p-1};n_{p-1})\right]$, where $\ba_i \in \bz_{q_1}\times\bz_{q_2}
\times\bz_{q_3}$, $n_i \in \bz_{q_1}$ and, among other things,
$\ba_i\neq \ba_j$ $(\forall i\neq j)$ holds.

Let us next discuss  the $(\bz_{q_1})^{q_1q_2q_3}$-invariance.   
Let $\bm \equiv \{m_{\bsa} \} \in (\bz_{q_1})^{q_1q_2q_3}$, 
and we define the action of $(\bz_{q_1})^{q_1q_2q_3}$ on the CP indices
of twisted sectors as follws;
\begin{equation}
\sigma_{\bsm} ({\cal A}) (\equiv (1,\bm)(\cA)) =  
\left[(\ba_0;n_0+m_{\bsa_0}),\ldots,(\ba_{p-1};n_{p-1}+m_{\bsa_{p-1}})\right],
\end{equation}
where ${\cal A}\equiv 
 \left[(\ba_0;n_0),\ldots , 
(\ba_{p-1};n_{p-1})\right]$ as above. 
This acts trivially on $X^{a\da}_{\cA}$ and
$\bar{\psi}^{\dal\da}_{\cA}$,
but acts on $\psi^{\al\da}_{\cA}$ in a non-trivial manner.  
To see it explicitly,
it is easier to consider the twisted fields 
$\{\psi^{\al\da \,(s)}_{\cA}\}_{s=0,\ldots, p-1}$ 
than $\psi^{\al\da}_{\cA}$ itself, and 
we can obtain from the condition \eqn{quiver psi}
\begin{equation}
\psi_{\sigma_{\bsm}(\cA)}^{\al\da\,(s)}
=\frac{1}{p}\sum_{s'=0}^{p-1}\sum_{r=0}^{p-1}\,\psi_{\cA}^{\al\da\,(s')}\,
   \om_p^{r(s'-s)}\om_{q_1}^{\al m_{\bsa_r}} .
\label{Z q1 psi}
\end{equation}
In the simplest example such that $m_{\bsa_0}=\cdots=m_{\bsa_{p-1}}=m$,
this reduces to 
\begin{equation}
 \psi^{\al \da}_{\sigma_{\bsm}({\cal A})}= \om^{m\al} \,
 \psi^{\al \da}_{{\cal A}} ,  \label{Z q1 psi 2} 
\end{equation}
and we also obtain for the NS vacuum $\ket{0,{\cal A}}$;
\begin{equation}
\begin{array}{l}
 \ket{0,\sigma_{\bsm}({\cal A})} = \ket{0,{\cal A}} ~~~(p=2k+1), \\
 \ket{0,\sigma_{\bsm}({\cal A})} = \om^{-m} \ket{0,{\cal A}} ~~~(p=2k).
\end{array}
\label{Z q1 psi 3}
\end{equation}
Recall that, when $p$ is odd,  $ \ket{0,{\cal A}}$ is 
$SU(2)_L$-neutral, but it has an extra $SU(2)_L$-charge 
$\dsp -\frac{1}{2}$ for the cases with even $p$.
The conditions \eqn{Z q1 psi 2}, \eqn{Z q1 psi 3} 
again imply that only the states possessing  
the $\om^{2J^3_0}$-invariance survive after performing the quiver
projection.

In summary  the chiral primaries of the twisted sectors in (0,4) theory
can be  constructed by the following procedures:
%(we shall again focus on only the ``surely stable'' excitations 
%as we already commented): 
\begin{itemize}
\item First, we should consider all the (single particle) chiral primaries
in the twisted sectors of the forms; 
$\cA= [(\ba_0;n_0),\ldots, (\ba_{p-1};n_{p-1})]$, $\ba_i\neq \ba_j 
~ (\forall i\neq j)$, which are essentially the twisted sectors 
in $Sym^{q_1q_2q_3}(M^4)$-$SCFT_2$ ({\em not\/} in   $Sym^{Q_1Q_5}(M^4)$). 
%%%
\item Second, consider $SU(1,1|2)_L$-module over each of them, 
and pick up the primary states such that 
\begin{equation}
\om^{2J^3_0} \ket{\al}=\ket{\al},
\end{equation}
which are the desired chiral primary states in our (0,4)-theory.
\end{itemize}
%The most important point is the following: The quiver projection breaks 
%the discrete symmetry from $S_{Q_1Q_5}$ to $S_{q_1q_2q_3}\times
%\bz_{q_1}$, and $\bz_{q_1}$-invariance on the CP indices becomes equivalent
%to the moding out by $\om^{2J^3_0}$. 
%after the condition 
%\eqn{quiver psi 2} is incorporated.

Since we already know the explicit forms of chiral primaries 
of (4,4)-theory, it is straightforward  to write down those for (0,4)-theory
according to the above prescription.
However, to avoid unessential complexity,
we here only present the spectrum of degeneracy of them with
certain  quantum numbers.
They correspond to the short representations 
of $SL(2,\br)_L\times SU(1,1|2)_R$, and 
we express each of them  by $(\bar{j};s)$, where
\begin{equation}
\begin{array}{l}
 \bar{\hat{J}}^3_0 \ket{\al} (\equiv \bar{\hat{L}}_0 \ket{\al} )
= \bar{j} \ket{\al}~~~(\mbox{R-charge}) \\
 (\hat{L}_0 -\bar{\hat{L}}_0 ) \ket{\al} = s \ket{\al}~~~(\mbox{spin})
\end{array}
\end{equation}
in the same way as in SUGRA.  
Moreover, we express the  space of $SL(2,\br)_L\times SU(1,1|2)_R$ irreps.
in which lowest component is a boson (fermion) by
${\cal H}^{(+)}_{\msc{SCFT}}$   
(${\cal H}^{(-)}_{\msc{SCFT}}$) 
as before.

Here we give a comment before we exhibit the spectrum of chiral primaries. 
By the above prescription we should take acount of  not only the states 
with $J^3_0=0$ but also with $\dsp J^3_0 = \frac{q_1 k}{2}$ 
$(\forall k \in \bz_{\geq 0})$. 
However, the states possessing non-zero $J^3_0$-charges  
correspond to  very massive modes in string theory. 
In fact, in the  original 
$AdS_3\times S^2$-geometry this charge is identified with the winding
along $x^5$-direction (in our T-dualized framework, the KK momentum
along the Taub-NUT circle), and hence the states with this winding 
are much heavier than the KK excitations on $S^3$
(We are now assuming these winding modes have the masses of the extent
of Planck mass, and the KK excitations on $S^3$ have the masses
$\dsp \sim \frac{1}{l^2}<<M_{\msc{Planck}}^2 $.),  which means 
these states are negligible in the region where SUGRA is valid.

Notice also that our quiver $SCFT_2$ describes the branes at 
the fixed point of the orbifold $\bc^2/\bz_{q_1}$, {\em which is the 
decompactification limit of the Taub-NUT space\/}. 
So, the stringy heavy states just mentioned become
lighter under this orbifold limit. 
The existence  of these extra states with non-zero $J^3_0$-charges
in our quiver $SCFT_2$ does not mean 
a failure of the $AdS_3/CFT_2$-correspondence.

Therefore, we here only present the spectra of the ``light'' 
chiral primaries with vanishing $J^3_0$-charges, which should be
compared with the results of SUGRA.
We will  give the complete spectra including non-zero $J^3_0$-charges
in Appendix, since they have rather involved forms.

By including   the contributions from  all of 
the untwisted and twisted sectors,
we can  obtain the following spectrum for $M^4=T^4$ 
(We set $D\equiv q_1q_2q_3$ again);
\begin{equation}
\begin{array}{l}
\dsp  {\cal H}^{(+)}_{\msc{SCFT}} (T^4) =  
(0 \,;\, 2)+5(0 \,;\,1)+(0\,;\,0)
+(1\,;\,2)+15(1\,;\,1)+14(1\,;\,0)+(1\,;\,-1) \\
  \dsp  ~~~~~ ~~~+\sum_{n=2}^{\left[\frac{D-1}{2}\right]}\, 
\left\{ (n \,;\,2)+15(n\,;\,1)+ 15(n\,;\,0)+ (n\,;\,-1)\right\}  \\
\dsp ~~~~~~~~ + \left\{
\begin{array}{ll}
  (N+1\,;1) + 6(N+1\,;0)+(N+1\,;-1) &(D=2N+1) \\
  9(N\,;1) + 14(N\,;0)+(N\,;-1) &(D=2N)
\end{array}\right.  \\
%%%
\dsp {\cal H}^{(-)}_{\msc{SCFT}} (T^4)= 
6(\frac{1}{2}\,;\,\frac{3}{2})+14(\frac{1}{2}\,;\,\frac{1}{2}) \\
\dsp ~~~~~ ~~~  +\sum_{n=2}^{\left[\frac{D-1}{2}\right]}\,
                \left\{6(n-\frac{1}{2}\,;\,\frac{3}{2}) 
         +20(n-\frac{1}{2}\,;\,\frac{1}{2})
        +6(n-\frac{1}{2}\,;\,-\frac{1}{2})\right\}  \\
\dsp ~~~~~~~~ + \left\{
\begin{array}{ll}
 \dsp 2(N+\frac{1}{2}\,;\frac{3}{2}) + 16(N+\frac{1}{2}\,;\frac{1}{2})
    + 6(N+\frac{1}{2}\,;-\frac{1}{2})  &(D=2N+1)\\
 \dsp 4(N+\frac{1}{2}\,;\frac{1}{2})
    + 4(N+\frac{1}{2}\,;-\frac{1}{2})  &(D=2N)
\end{array} \right. 
\end{array} 
\label{CP T6} 
\end{equation}
It is worth mentioning  that the R-charge in this spectrum
has upper bound $\dsp \sim \frac{D}{2}$. 
This  is because the length of maximally twisted string becomes
$D\equiv q_1q_2q_3$ ({\em not\/} $Q_1Q_5$) after performing the 
quiver projection.

~

%%%%%%%%%%%%%%%%%%%%%%%%%%%%%%%%%%%%%%%%%%%%%%%%%%%%%%%%%%%%%%%%%%%%%%%%%%
Let us now turn to the studies  for $M^4=K3$.
As in the previous discussions in the untwisted sector, 
we regard as $K3 \cong T^4/\bz_2$. 

For the untwisted sector in the sense of  $\bz_2$-orbifold,
we need to make little change in the above results for $M^4=T^4$.
The requirement of $\bz_2$-invariance simply kills the odd cohomologies 
of $T^4$, and we obtain the contributions from the $(4,4)$
chiral primaries of the following types;
$\ket{\om^{(0,0)},(p-1)}$, $\ket{\om^{(2,0)},(p-1)}$, 
$\ket{\om^{(0,2)},(p-1)}$, $\ket{\om^{(2,2)},(p-1)}$ 
and also the four series  of the type  $\ket{\om^{(1,1)},(p-1)}$. 

To get the complete spectrum we must also make the twisted sectors of 
$\bz_2$-orbifold  join in our game. There are 16 twisted sectors 
corresponding to each fixed points of $T^4/\bz_2$. 
We here only focus on one of them, because all of them have the same 
algebraic structure. At the same time we take the $\bz_p$-twisted sector 
in the sense of the symmetric orbifold.
\begin{itemize}
 \item $p=2k+1$

We have the next mode expansions for the bosonic and fermionic oscillators;
$\al^{a\da}_{{\cal A}, \, n+\frac{1}{2}}$, 
$\psi^{\al\da}_{{\cal A}, \, n}$, $(n\in \bz)$,
and for the NS vacuum $\ket{0,{\cal A}}$ (We omit the label which 
distinguishes the 16 sectors for simplicity.) we have
\begin{equation}
\begin{array}{l}
\dsp \hat{L_0}\ket{0,{\cal A}} = \left(\frac{p^2-1}{4p} + \frac{1}{2p}
\right)\ket{0,{\cal A}} \equiv \frac{p^2+1}{4p}\ket{0,{\cal A}}, \\
\dsp \hat{J}_0^3\ket{0,{\cal A}}= -\frac{1}{2}\ket{0,{\cal A}}.
\end{array}
\label{K3 twisted odd}
\end{equation}
\item $p=2k$

We have the mode expansions
$\al^{a\da}_{{\cal A}, \, n+\frac{1}{2}}$, 
$\psi^{\al\da}_{{\cal A}, \, n+\frac{1}{2}}$, $(n\in \bz)$,
and for the NS vacuum
\begin{equation}
\begin{array}{l}
\dsp \hat{L_0}\ket{0,{\cal A}} = \left(\frac{p^2-1}{4p} + \frac{1}{4p}
\right)\ket{0,{\cal A}} \equiv \frac{p}{4}\ket{0,{\cal A}}, \\
\dsp \hat{J}_0^3\ket{0,{\cal A}}= 0 .
\end{array}
\label{K3 twisted even}
\end{equation}
\end{itemize}
In these expressions 
the extra zero-point energies and R-charges assigned to the NS vacuua
have their origins in these twisted mode expansions. 
%of the bosonic and fermionic oscillators. 
The superconformal currents 
in this twisted sector (that is, the ``twisted''  in the meaning of   
both $T^4/\bz_2$ and the symmetric orbifold) are also defined
as before \eqn{hat}, and they give us 
only one chiral primary state for each sector,   which has 
the form $\ket{\om^{(1,1)},(p-1)}$;
\begin{equation}
\ket{\om^{(1,1)},(p-1)}=\sum_{{\cal A}}\,\ket{\om^1,{\cal A}}\otimes
   \overline{\ket{\om^1,{\cal A}}} ,
\label{CP K3 twisted 1}
\end{equation}
\begin{equation}
\begin{array}{l}
\dsp  \ket{\om^1,{\cal A}}= \prod_{i=0}^{k-1}\,
\psi^{++}_{{\cal A}\, -i}
\psi^{+-}_{{\cal A}\, -i} \ket{0,{\cal A}}, ~~~(p=2k+1)\\
\dsp  \ket{\om^1,{\cal A}}= \prod_{i=0}^{k}\,
\psi^{++}_{{\cal A}\, -\left(\frac{1}{2}+i\right)}
\psi^{+-}_{{\cal A}\, -\left(\frac{1}{2}+i\right)} 
\ket{0,{\cal A}}. ~~~(p=2k)
\end{array}
\label{CP K3 twisted 2}
\end{equation}
We can easily check that they have the expected
quantum numbers
\begin{equation}
\hat{L}_0 \ket{\om^1,{\cal A}} = \hat{J}^3_0\ket{\om^1,{\cal A}}
= \frac{p}{2} \ket{\om^1,{\cal A}}.
\end{equation}
Of course, there are 16 independent such  states 
corresponding to each twisted sectors of $T^4/\bz_2$, 
and they compensate the missing
cohomologies of $Sym^{Q_1Q_5}(K3)$. Thus we can get the 20
sequences of the form $\ket{\om^{(1,1)},(p-1)}$ 
by including also  the 4 contributions from the untwisted sector.

The next step is the same as that in the case of $T^4$.
We can take the quiver projection for this twisted sector
and can obtain the spectrum which is modded out by $\om^{2J_0^3}$.
Gathering the contributions from all the sectors, we can 
lastly obtain the next spectrum for $K3$;
\begin{equation}
\begin{array}{l}
\dsp  {\cal H}^{(+)}_{\msc{SCFT}} (K3) =  
(0 \,;\, 2)+ (0 \,;\,1)+(0\,;\,0) 
+(1\,;\,2)+23(1\,;\,1)+22(1\,;\,0)+(1\,;\,-1) \\
  \dsp  ~~~~~ ~~~+\sum_{n=2}^{\left[\frac{D-1}{2}\right]}\, 
\left\{ (n \,;\,2)+23(n\,;\,1)+ 23(n\,;\,0)+ (n\,;\,-1)\right\}  \\
\dsp ~~~~~~~~ + \left\{
\begin{array}{ll}
  (N+1\,;1) + 2(N+1\,;0)+(N+1\,;-1) &(D=2N+1) \\
  21(N\,;1) + 22(N\,;0)+(N\,;-1) &(D=2N)
\end{array}\right. , \\
%%%
\dsp {\cal H}^{(-)}_{\msc{SCFT}} (K3)= 
2(\frac{1}{2}\,;\,\frac{3}{2})+42(\frac{1}{2}\,;\,\frac{1}{2}) \\
\dsp ~~~~~ ~~~  +\sum_{n=2}^{\left[\frac{D-1}{2}\right]}\,
                \left\{2(n-\frac{1}{2}\,;\,\frac{3}{2}) 
         +44(n-\frac{1}{2}\,;\,\frac{1}{2})
        +2(n-\frac{1}{2}\,;\,-\frac{1}{2})\right\}  \\
\dsp ~~~~~~~~ + \left\{
\begin{array}{ll}
 \dsp  42(N+\frac{1}{2}\,;\frac{1}{2})
    + 2(N+\frac{1}{2}\,;-\frac{1}{2})  &(D=2N+1)\\
 \dsp  2(N+\frac{1}{2}\,;\frac{1}{2})
    + 4(N+\frac{1}{2}\,;-\frac{1}{2})  &(D=2N)
\end{array} \right. .
\end{array} 
\label{CP K3} 
\end{equation}

~

%%%%%%%%%%%%%%%%%%%%%%%%%%%%%%%%%%%%%%%%%%%%%%%%%%%%%%%%%%%%%%%%%%%%%%%%%%%%%
To end this subsection we present a comparison with the results of SUGRA
given in the previous section.
By comparing the results \eqn{KK T6}, \eqn{KK K3} and \eqn{CP T6}, 
\eqn{CP K3}, we can notice that both are equivalent except the following
slight differences: 

First, we now  have  the upper bound for R-charges, which is 
essentially the one first remarked in \cite{MS} (``stringy exclusion 
principle''). This difference is not problematic at all,  as far as we work 
in the  sufficiently large $D\equiv q_1q_2q_3$ assumed in the Maldacena
conjecture.  

The second difference is owing to the existence of the chiral primaries 
$\ket{\om^{(0,2)},(0)}$, $\ket{\om^{(2,0)}, (0)}$ in the (4,4) theory.
They correspond to the degrees of freedom which essentially exists
only on the boundary (called ``singleton'')
\cite{Larsen,deBoer1}. Thus we have no contradiction even if we cannot 
find them in the bulk theory.
 
In this way we can conclude that the spectrum of chiral primaries
of our quiver $SCFT_2$ is consistent with that of the SUGRA in 
the $AdS_3\times S^2$-geometry at least for the single particle states.  
We must further study the multi-particle states to claim 
the complete agreement between them. 
The next subsection will devote to this subject.

~

%%%%%%%%%%%%%%%%%%%%%%%%%%%%%%%%%%%%%%%%%%%%%%%%%%%%%%%%%%%%%%%%%%%%%%%%%
\subsection{Discussions about Multi-Paritcle States}

%\hspace*{4.5mm}

In the previous subsection we have shown that our quiver $SCFT_2$ can 
produce the spectrum of single particle states which is consistent with that
of SUGRA on $AdS_3\times S^2$.  The essential part of our discussion 
was  that the quiver projection works on the Hilbert space
as moding out by $\om^{2J^3_0}$. At this point one might ask
the followin questions:

{\em Is the quiver projection truly needed? 
Why can't we define the Hilbert space of the boundary (0,4) $SCFT_2$
by directly imposing the constraints $\om^{2J^3_0}\ket{\al}=\ket{\al}$ 
on the Hilbert space of (4,4) $SCFT_2$ on $Sym^{Q_1Q_5}(M^4)$?}
 
The answer to the first question is {\em Yes}, and we can answer
to the second question as follows: 
One of the evidences for us to need the quiver projection
is  given by evaluating the  degrees of freedom. If we naively mode out  
by $\om^{2J^3_0}$ to define the (0,4) theory (To get  a  more suitable 
candidate of the  (0,4) theory, 
one should further incorporate the twisted sectors of 
this orbifoldization to ensure the modular invariance), 
we will obtain the (0,4) $SCFT_2$ with 
the same central charge $c=6Q_1Q_5$ as that of (4,4) theory.
But we want the $SCFT_2$ with the $1/q_1$ times degrees of freedom. 
As was already seen, the quiver projection
actually yields the (0,4) $SCFT_2$ with the central charge
$\dsp c=\frac{6Q_1Q_5}{q_1}\equiv 6q_1q_2q_3$.   

The second reason why we need to consider the quiver projection 
is more crucial and has a deep relationship 
to the analysis of the multi-particle states. 
Roughly speaking, in order to be consistent with the spectrum of SUGRA, 
the multi-particle states in the boundary $SCFT_2$
should have the structure
$\sim \cO_1\cO_2\cdots \ket{0}$, where $\cO_i$ denotes some descendant
of the chiral primary of single particle type.
As was already mentioned, in \cite{deBoer2,MMS}
it is shown that, in the (4,4) $SCFT_2$ on $Sym^{Q_1Q_5}(M^4)$,
general BPS states which contribute to the invariant index 
(elliptic genus for $K3$, the index given in \cite{MMS} for $T^4$)
should have indeed such forms, if their R-charges are less than
$\dsp \frac{Q_1Q_5+1}{4}\sim \frac{Q_1Q_5}{4} $. In other words
the correspondence of the chiral primaries between
the $AdS_3\times S^3$-SUGRA and the boundary (4,4) $SCFT_2$ 
are satisfactory in the level of multi-particle states
as far as taking the large $Q_1Q_5$-limit.
%\footnote
%     {The analysis by means of elliptic genus
%      given in \cite{deBoer2} has no power in  the case of $M^4=T^4$, 
%     because the elliptic genus is trivially zero for the toroidal model. 
%     After I submitted the first version of this paper, it was proved
%      \cite{MMS} that there is also the similar good correspondence 
%     below the R-charge bound $\dsp \sim \frac{Q_1Q_5}{4}$ 
%     for $M^4=T^4$, based on the analysis of 
%     the ``new SUSY index''.  Hence the following arguments 
%     can be safely applied to the case $M^4=T^4$, too.}. 

Now, we first try to define 
the Hilbert space of the (0,4) theory,
which we express by ${\cal H}^{(0,4)'}$ in order to distinguish
it from ${\cal H}^{(0,4)}$ \eqn{(0,4) Hilbert space} defined by the 
quiver projection,
by simply imposing the constraints $\om^{2J^3_0}\ket{\al}=\ket{\al}$ on
the Hilbert space of (4,4) theory  ${\cal H}^{(4,4)}$ 
%in place of 
%taking the quiver projection discussed in the previous section. 
(Here, we overlook the problem about the central charge we already mentioned.)
As we observed before, 
the ${\cal H}^{(0,4)'}$ so defined includes all the single particle states
with correct quantum numbers.
But, how about the multi-particles? 
If we admit the complete agreement between the spectrum of (4,4) $SCFT_2$
and the $AdS_3\times S^3$-SUGRA, 
and adopt ${\cal H}^{(0,4)'}$ as the Hilbert space
of (0,4)-theory,  we will obtain the following general
forms of the BPS states of (0,4) $SCFT_2$;
\begin{equation}
 \ket{\al}= \cO_1\cO_2\cdots \ket{0}, ~~~[2J_0^3,\, \cO_i\} = r_i\cO_i ,~~~
\sum_{i}r_i \in q_1\bz,
\label{multi 1}
\end{equation}   
where $\cO_i$ denotes the $1/4$ BPS observables of the single particle
type in the sense of (4,4) SUSY, and 
the reader should read $[~,~\}$ as the (anti-)commutator
for the bosonic (fermionic) field $\cO_i$.
The important point is that each $r_i$ need not be a multiple of $q_1$,
although $\ket{\al}$ must be  $\bz_{q_1}$-invariant.  
Such $\ket{\al}$ is of course a multi-particle states in the sense 
of the (4,4) theory, but generally  {\em not\/} in the (0,4) theory, 
because the constituents $\cO_i$
may not belong to the Hilbert space ${\cal H}^{(0,4)'}$. 

On the other hand, the Hilbert space of multi-particle states 
in SUGRA should be defind by taking the (symmetric or anti-symmetric)
tensor products of single particle states. This fact brings us the result
that  the SUGRA on $AdS_3\times S^3/\bz_{q_1}$ should
include only the multi-particle states of the following type; 
\begin{equation}
 \ket{\al}= \cO_1\cO_2\cdots \ket{0}, ~~~[2J_0^3,\, \cO_i\} = r_i\cO_i ,~~~
 (\forall r_i \in q_1\bz).
\label{multi 2}
\end{equation}   
Clearly the states with the form \eqn{multi 2} merely spans  
the subspace of that for  \eqn{multi 1}. We thus face  the existence of
many missing states, if we adopt the simple prescription  of 
orbifoldization by $\om^{2J^3_0}$!

Next let us observe how we can resolve this inconsistency by considering 
the quiver projection.
We should notice that each of single-paricle operators has the next 
CP index structure;
$\dsp \cO_i = \sum_{{\cal A}^{(i)}} \cO_{i,\,{\cal A}^{(i)}}$, where
${\cal A}^{(i)}\equiv \left[(\ba_0^{(i)};n_0^{(i)}),\ldots
(\ba_{p_i-1}^{(i)};n_{p_i-1}^{(i)})\right]$ 
labels one of the $\bz_{p_i}$-twisted sector. (If $p_i=1$, one 
should understand  it as the untwisted sector, of course.)
The general multi-particles in the (4,4) theory have the forms;
\begin{equation}
\ket{\al}= \cO_1\cO_2 \cdots \ket{0}\equiv
\left(\sum_{{\cal A}^{(1)}} \cO_{1,\,{\cal A}^{(1)}}\right)\,
\left(\sum_{{\cal A}^{(2)}} \cO_{2,\,{\cal A}^{(2)}}\right)\, \cdots \ket{0} .
\end{equation}

The quiver projection again brings us the requirements of the 
$(\bz_{q_1})^{q_1q_2q_3}$-invariance with respect to the CP indices.
Especially, we can take  an element 
$\bm \equiv \{m_{\bsa}\}$ of $(\bz_{q_1})^{q_1q_2q_3}$   
such that $m_0^{(i)}=\cdots=m_{p_i-1}^{(i)}=m^{(i)}$, 
where $m_r^{(i)}\equiv m_{\bsa_r^{(i)}}$.
It is the most important to notice  that {\em we can  choose
these $m^{(i)} \in \bz_{q_1} $ 
independently for each $i$, because of the ``locality''
of the $(\bz_{q_1})^{q_1q_2q_3}$-symmetry.}
Therefore, according to the discussions given  in the previous 
subsections, we find that  the $\om^{2J^3_0}$-invariance 
are {\em separately\/} imposed on each of the single particle components
$\cO_i$,  in contrast to the usual orbifoldization of CFT. 
In other words, ${\cal H}^{(0,4)}$ only  includes the BPS states  
possessing  the forms \eqn{multi 2}, 
although ${\cal H}^{(0,4)'}$ also includes those of the forms \eqn{multi 1}.

In this way we can conclude  that 
the quiver projection does reproduce  
the consistent spectrum of multi-particles \eqn{multi 2}.   
We can also rephrase: {\em The orbifoldization of bulk
SUGRA does not correspond to the orbifoldization of the boundary CFT,
rather correspond to the quiver projection of it!}

~

\section{Summary and Comments}

\cleqn
\hspace*{4.5mm}

In this paper we have proposed a candidate for 
the suitable boundary $SCFT_2$ which is dual to 
the SUGRA on $AdS_3\times S^2 \times M^6$ $(M^6=T^6,~T^2\times K3)$ 
by making use of the quiver techniques. 
We presented a detailed analysis on the chiral 
primaries, including all of the twisted sectors of symmetric orbifolds, 
and obtained the spectrum which is consistent with SUGRA
in the Maldacena limit.  

It is plausible and may not be  so surprising that the quiver projection 
on the single particle states leads to the simple 
moding out prescription by $\om^{2J^3_0}$. 
But for the multi-particle states, this  procedure of orbifolding 
leads to the wrong results. We must carefully consider the quiver
projection, not the orbifolding, to get a consistent spectrum of 
the multi-particles.

However, because of just  this fact that our quiver $SCFT_2$
is not an usual orbifold $SCFT_2$, it will become a subtle problem whether
it has actually the modular invariance.  
Needless to say, the modular invariance
is an important consistency condition of $CFT_2$ on torus, and in addition,
in the thermodynamical studies on the $AdS_3$-gravity including 
some BTZ black-holes \cite{BTZ}, it is expected to yield  a very powerful  
tool \cite{Banados,MS}. 
The check of modular invariance for the quiver $SCFT_2$  
seems a challenging and an important  problem, and we hope to 
investigate this problem in future.

One of the natural generalizaitons of the studies in this paper is 
the works for $M^6=CY_3$. Although the analyses for the completely
general $CY_3$ seem very difficult, it might be possible to apply 
our formulation at least to some elliptic $CY_3$.
We can reduce by the ``fiber-wise T-duality'' the original system
of M5 to the one composed of a D3 wrapped around a SUSY 2-cycle
in the base manifold of the elliptic fibration, 
located at the singularity of an orbifold. 
%This is expected 
%to be described by the $N=(4,4)$ $\sigma$-model on some Hitchin space
%(see for example \cite{BSV}) with SUSY broken by the quiver projection.

It may be also interesting to compare our  quiver $SCFT_2$ with 
the ``space-time CFT'' \cite{KLL} given 
in the string theoretical approach along the line of \cite{GKS}.
It is worth remarking that the quiver projection 
can completely reproduce the (0,4) version of the 
``stringy exclusion principle'' of \cite{MS} (That is,
the unitarity bound of the arbitrary chiral primary is 
$\dsp j\leq q_1q_2q_3 \left(\equiv \frac{Q_1Q_5}{q_1}\right) $, and 
that for the single particle is 
$\dsp j\leq \frac{q_1q_2q_3}{2} $. The latter was already mentioned, and 
the former is easily shown by counting the degrees of freedom and 
by taking acount of the fermi statistics.).  
Whether this bound can be also  explained in the framework of \cite{KLL}
is  an important test. To answer this question completely,  
we may need first to study carefully the unitarity bound for the (4,4)
space-time CFT \cite{GKS}, maybe based on the discussions given in
\cite{EGP}. 
To this aim the approach from the Matrix string theory \cite{Matrix}
may become more efficient  than 
that  of the ``old fashioned'' string on $AdS_3$ back-ground.
Some studies along this line will be presented in \cite{HS}.

~

\section*{Acknowledgement}

\hspace*{4.5mm}

I would like to thank T. Eguchi for valuable discussions,
and especially should appriciate T. Kawano and K. Okuyama
for their collaboration in the early stage of this work. 

This work is partly supported  by the Grant-in-Aid
for Scientific Research on Priority Area 707 
``Supersymmetry and Unified Theory
of Elementary Particles", Japan Ministry of Education.

%%%%%%%%%%%%%%%%%%%%%%%%%%%%%%%%%%%%%%%%%%%%%%%%%%%%%%%%%%%%%%%%%%%%%%%%%%
\newpage
\noindent
{\Large \bf Appendix}
\appendix
\section{Complete Spectra of the Chiral Primaries in Quiver $SCFT_2$}

\cleqn
\hspace*{4.5mm}

Here we give the complete spectra of 
the chiral primaries including the general states such 
that $2J^3_0\ket{\al}=k q_1\ket{\al}$ $(k\in \bz_{\geq 0})$.
To this aim it is convenient to introduce the symbol $(\bar{j};s)_k$
for the irreducible $SL(2,\br)_L\times SU(1,1|2)_R$-modules
in which all the states  should satisfy 
\begin{equation}
2J^3_0 \ket{\al} =k q_1\ket{\al},
\end{equation}
and the highest weight state (chiral primary) does
\begin{equation}
\bar{L}_0 \ket{\al}=\bar{J}^3_0\ket{\al} =(\bar{j}+\frac{kq_1}{2})\ket{\al},
~~~(L_0-\bar{L}_0)\ket{\al}=s\ket{\al}.
\end{equation}
Under this preparation we can write down the desired  spectra of 
chiral primaries as follows ($D\equiv q_1q_2q_3$ as before);
\begin{itemize}
\item $M^4=T^4$
\begin{equation}
\begin{array}{lll}
{\cal H}^{(+)}_{\msc{SCFT}}(T^4) &= &
\dsp \sum_{k=0}^{q_2q_3-1} \,\left\{
(0\,;2)_k+5(0\,;1)_k+(0\,;0)_k+(1\,;2)_k +6(1\,;1)_k+(1\,;0)_k \right\} \\
& & \dsp + \sum_{k=0}^{q_2q_3} \,\left\{
9(1\,;1)_k+13(1\,;0)_k+(1\,;-1)_k
\right\} \\
&&\dsp +\sum_{n=2}^{\left[\frac{D-1}{2}\right]}
        \sum_{k=0}^{\left[\frac{D-2n}{q_1}\right]}\,
\left\{(n\,;2)_k+15(n\,;1)_k +15(n\,;0)_k+(n\,;-1)_k \right\} \\
&&\dsp + \left\{
\begin{array}{ll}
 (N+1\,;1)+6(N+1\,;0)+(N+1\,;-1) & (D=2N+1) \\
 9(N\,;1)+14(N\,;0)+(N\,;-1) & (D=2N)
\end{array}
\right.
\end{array} 
\end{equation}
%%%
\begin{equation}
\begin{array}{lll}
{\cal H}^{(-)}_{\msc{SCFT}}(T^4) &= &
\dsp \sum_{k=0}^{q_2q_3-1} \,\left\{4(\frac{1}{2}\,;\frac{3}{2})_k 
+4(\frac{1}{2}\,;\frac{1}{2})_k  \right\} 
+ \sum_{k=0}^{q_2q_3} \,\left\{
2(\frac{1}{2}\,;\frac{3}{2})_k+10(\frac{1}{2}\,;\frac{1}{2})_k
\right\} \\
&&\dsp +\sum_{n=2}^{\left[\frac{D-1}{2}\right]}
        \sum_{k=0}^{\left[\frac{D-2n}{q_1}\right]}\,
\left\{6(n-\frac{1}{2}\,;\frac{3}{2})_k
+20(n-\frac{1}{2}\,;\frac{1}{2})_k
+6(n-\frac{1}{2}\,;-\frac{1}{2})_k \right\} \\
&&\dsp + \left\{
\begin{array}{ll} 
\dsp 2(N+\frac{1}{2}\,;\frac{3}{2})+16(N+\frac{1}{2}\,;\frac{1}{2})
  +6(N+\frac{1}{2}\,;-\frac{1}{2}) & (D=2N+1) \\
\dsp 4(N+\frac{1}{2}\,;\frac{1}{2})
  +4(N+\frac{1}{2}\,;-\frac{1}{2}) & (D=2N)
\end{array}
\right.
\end{array}
\end{equation}
%%%%%
\item $M^4=K3$
\begin{equation}
\begin{array}{lll}
{\cal H}^{(+)}_{\msc{SCFT}}(K3) &= &
\dsp \sum_{k=0}^{q_2q_3-1} \,\left\{(0\,;2)_k 
+(0\,;1)_k + (0\,;0)_k+ (1\,;2)_k +2 (1\,;1)_k +  (1\,;0)_k \right\} \\
& & \dsp + \sum_{k=0}^{q_2q_3} \,\left\{
21(1\,;1)_k+21(1\,;0)_k+(1\,;-1)_k
\right\} \\
&&\dsp +\sum_{n=2}^{\left[\frac{D-1}{2}\right]}
        \sum_{k=0}^{\left[\frac{D-2n}{q_1}\right]}\,
\left\{(n\,;2)_k+23(n\,;1)_k +23(n\,;0)_k+(n\,;-1)_k \right\} \\
&&\dsp + \left\{
\begin{array}{ll}
 (N+1\,;1)+2(N+1\,;0)+(N+1\,;-1) & (D=2N+1) \\
 21(N\,;1)+22(N\,;0)+(N\,;-1) & (D=2N)
\end{array}
\right.
\end{array} 
\end{equation}
%%%
\begin{equation}
\begin{array}{lll}
{\cal H}^{(-)}_{\msc{SCFT}}(K3) &= &
\dsp \sum_{k=0}^{q_2q_3-1} \,\left\{2(\frac{1}{2}\,;\frac{3}{2})_k 
+2(\frac{1}{2}\,;\frac{1}{2})_k  \right\} 
+ \sum_{k=0}^{q_2q_3} \, 
40(\frac{1}{2}\,;\frac{1}{2})_k  \\
&&\dsp +\sum_{n=2}^{\left[\frac{D-1}{2}\right]}
        \sum_{k=0}^{\left[\frac{D-2n}{q_1}\right]}\,
\left\{2(n-\frac{1}{2}\,;\frac{3}{2})_k
+44(n-\frac{1}{2}\,;\frac{1}{2})_k
+2(n-\frac{1}{2}\,;-\frac{1}{2})_k \right\} \\
&&\dsp + \left\{
\begin{array}{ll} 
\dsp 42(N+\frac{1}{2}\,;\frac{1}{2})
 + 2(N+\frac{1}{2}\,;-\frac{1}{2}) & (D=2N+1) \\
\dsp 2(N+\frac{1}{2}\,;\frac{1}{2})
 + 2(N+\frac{1}{2}\,;-\frac{1}{2}) & (D=2N)
\end{array}
\right.
\end{array}
\end{equation}
\end{itemize}

%%%%%%%%%%%%%%%%%%%%%%%%%%%%%%%%%%%%%%%%%%%%%%%%%%%%%%%%%%%%%%%%%%%%%%%%%%


\newpage

\begin{thebibliography}{99}

\bibitem{Mal}
J. Maldacena,
%{\it``The Large $N$ Limit of Superconformal Field Theories
%and Supergravity''},
Adv. Theor. Math. Phys. {\bf 2} (1998) 231-252,
hep-th/9711200.

\bibitem{Witten1}
E. Witten,
%{\it ``Anti De Sitter Space And Holography''},
Adv. Theor. Math. Phys. {\bf 2} (1998) 253-291,
hep-th/9802150.

\bibitem{GKP}
S. Gubser, I. Klebanov and A. Polyakov,
%{\it ``Gauge Theory Correlators from Non-Critical String Theory''},
Phys. Lett. {\bf B428} (1998) 105-114,
hep-th/9802109.


\bibitem{Banados}
M. Banados, T. Brotz and M. Ortiz,
%{\it ``Boundary dynamics and the statistical mechanics of
%the 2+1 dimensional black hole''},
hep-th/9802076; 
M. Banados,
hep-th/9901148, and references therein.
%{\it ``Global Charges in Chern-Simons theory and the $2+1$ black hole''},
%Phys. Rev. {\bf D52} (1996) 5816,
%hep-th/9405171;
%\bibitem{9805165}
%M. Banados, K. Bautier, O. Coussaert, M. Henneaux, M. Ortiz,
%{\it ``Anti-de Sitter/CFT Correspondence in Three-Dimensional Supergravity''},
%Phys. Rev. {\bf D58} (1998) 085020,
%hep-th/9805165.


\bibitem{Martinec}
E. Martinec,
%{\it ``Matrix Models of $AdS$ Gravity''},
hep-th/9804111, 
%{\it ``Conformal Field Theory, Geometry, and Entropy''},
hep-th/9809021

\bibitem{deBoer1}
J. de Boer,
%{\it ``Six-Dimensional Supergravity on $S^3 \times AdS_3$
% and $2$d Conformal Field Theory''},
hep-th/9806104.

\bibitem{deBoer2}
J. de Boer,
%{\it ``Large $N$ Elliptic Genus and AdS/CFT Correspondence'''},
hep-th/9812240

\bibitem{MMS}
J. Maldacena, G. Moore and A. Strominger,
hep-th/9903163.


\bibitem{MSW}
J. Maldacena, A. Strominger and E. Witten,
%{\it ``Black Hole Entropy in M-Theory''},
JHEP {\bf 9712} (1997) 002,
hep-th/9711053.

\bibitem{Vafa}
C. Vafa,
%{\it ``Black Holes and Calabi-Yau Threefolds''},
Adv. Theor. Math. Phys. {\bf 2} (1998) 207-218,
hep-th/9711067.

\bibitem{BBG}
K. Behrndt, Ilka Brunner and I. Gaida,
%{\it ``$AdS_3$ Gravity and Conformal Field Theories''},
hep-th/9806195.


\bibitem{KLL}
D. Kutasov, F. Larsen, and R. Leigh,
%{\it ``String Theory in Magnetic Monopole Backgrounds''},
hep-th/9812027.

\bibitem{BL}
D. Berenstein and R. Leigh,
hep-th/9812142.

\bibitem{DM}
M. R. Douglas and G. Moore, 
%{\it ``D-branes, Quivers, and ALE Instantons''},
hep-th/9603167.


\bibitem{harmo}
A.A. Tseytlin,
%{\it ``Harmonic superpositions of M-branes''},
Nucl.Phys. {\bf B475} (1996) 149-163,
hep-th/9604035;
I.R. Klebanov and A.A. Tseytlin,
%{\it ``Intersecting M-branes as Four-Dimensional Black Holes''},
Nucl.Phys. {\bf B475} (1996) 179-192,
hep-th/9604166.


\bibitem{BH}
J. Brown, M. Henneaux, 
%{\it ``Central Charges In The Canonical Realization Of Asymptotic Symmetries:
%An Example from Three-Dimensional Gravity''},
Commun. Math. Phys. {\bf 104} (1986) 207.

\bibitem{Stro}
A. Strominger,
%{\it ``Black Hole Entropy from Near-Horizon Microstates''},
JHEP {\bf 9802} (1998) 009,
hep-th/9712251.

\bibitem{GKS}
A. Giveon, D. Kutasov and N. Seiberg,
%{\it ``Comments on String Theory on $AdS_3$''},
Adv. Theor. Math. Phys. {\bf 2} (1998) 733-780,
hep-th/9806194.

\bibitem{KS}
S. Kachru and E. Silverstein,
%{\it ``4d Conformal Field Theories and Strings on Orbifolds''},
Phys.Rev.Lett. {\bf 80}  (1998) 4855-4858,
hep-th/9802183.

\bibitem{LNV}
A. Lawrence, N. Nekrasov and C. Vafa,
%{\it ``On Conformal Field Theories in Four Dimensions''},
Nucl.Phys. {\bf B533} (1998) 199-209,
hep-th/9803015. 

\bibitem{OT}
Y. Oz and J. Terning,
%{\it ``Orbifolds of $AdS_5\times S^5$ and $4d$ Conformal Field
% Theories''},
Nucl.Phys. {\bf B532} (1998) 163-180,
hep-th/9803167.

\bibitem{Gukov}
S. Gukov,
%{\it``Comments on $N=2$ $AdS$ Orbifolds''},
Phys. Lett. {\bf B439} (1998) 23-28,
hep-th/9806180;
S. Gukov, M. Rangamani and E. Witten,
%{\it ``Dibaryons, Strings, and Branes in $AdS$ Orbifold Models''},
hep-th/9811048.


\bibitem{junction}
K. Dasgupta and S. Mukhi,
Phys. Lett. {\bf B423} (1998) 261, hep-th/9711094;
C. Callan and L. Thorlacius,
Nucl. Phys. {\bf B534} (1998) 121, hep-th/9803097.

\bibitem{Larsen}
F. Larsen,
%{\it ``The Perturbation Spectrum of Black Holes in $N=8$
% Supergravity''},
Nucl.Phys. {\bf B536} (1998) 258-278,
hep-th/9805208.

\bibitem{FKM}
A. Fujii, R. Kemmoku and S. Mizoguchi,
%{\it ``$D=5$ Simple Supergravity on $AdS_3\times S^2$ and $N=4$
%Superconformal Field Theory''},
hep-th/9811147.

\bibitem{KN}
P. Kronheimer and H. Nakajima,
%{\it ``Yang-Mills instantons on ALE gravitational instatantons''},
Math. Ann. {\bf 288} (1990) 263.


\bibitem{Witten2}
E. Witten,
%{\it ``Sigma Models And The ADHM Construciton Of Instantons''},
J. Geom. Phys. {\bf 15}  (1995) 215,
hep-th/9410052.

\bibitem{Lambert}
N. Lambert,
%{\it ``D-brane Bound States and the Generalised ADHM Construction''},
Nucl.Phys. {\bf B519} (1998) 214-224,
hep-th/9707156.


\bibitem{HW}
S. Hassan and S. Wadia,
%{\it ``Gauge Theory Description of D-brane Black Holes: Emergence of the 
%Effective SCFT and Hawking''},
Nucl. Phys. {\bf B526}(1998) 311, 
hep-th/9712213.


\bibitem{MS}
J. Maldacena, A. Strominger,
%{\it ``$AdS_3$ Black Holes and a Stringy Exclusion Principle''},
JHEP {\bf 9812} (1998) 005,
hep-th/9804085.

\bibitem{Dijkgraaf}
R. Dijkgraaf,
%{\it ``Instanton Strings and HyperK\"{a}hler Geometry''},
hep-th/9810210.

\bibitem{orbifold}
L. Dixon, J. Harvey, C. Vafa and E. Witten,
Nucl. Phys. {\bf B261} (1985) 678; 
L. Dixon, D. Friedan, E. Martinec and E. Witten,
Nucl. Phys. {\bf B281} (1987) 13.

\bibitem{DMVV}
R. Dijigraaf, G. Moore, E. Verlinde and H. Verlinde,
Commun. Math. Phys. 185 (1997) 197.

\bibitem{Witten3}
E. Witten,
%{\it ``On the Conformal Field Theory of the Higgs Branch''},
JHEP {\bf 9707} (1997) 003,
hep-th/9707093.

\bibitem{VW}
C. Vafa and E. Witten, 
Nucl. Phys. {\bf B431} (1994), 3, and references therein.

\bibitem{KSSW}
D. Kutasov and N. Seiberg,
%{\it ``More Comments on String Theory on $AdS_3$''},
hep-th/9903219;
N. Seiberg and E. Witten,
%{\it ``The D1/D5 System and Singular CFT''},
hep-th/9903224.

\bibitem{HS}
K. Hosomichi and Y. Sugawara,
JHEP {\bf 9901} (1999) 013, hep-th/9812100;
K. Hosomichi and Y. Sugawara, 
hep-th/9905004


\bibitem{FKN}
M. Fukuma, H. Kawai and R. Nakayama,
Comm. Math. Phys. {\bf 143} 371-403 (1992).

\bibitem{BHS}
L. Borisov, M. Halpern and C. Schweigert,
%{\it ``Systematic Approach to Cyclic Orbifolds''},
Int. J. Mod. Phys. {\bf A13} (1998) 125-168
hep-th/9701061.

\bibitem{BTZ}
M. Banados, C. Teitelboim, J. Zanelli,
%{\it ``The Black Hole in Three Dimensional Space Time''},
Phys. Rev. Lett. {\bf 69} (1992) 1849,
hep-th/9204099.

\bibitem{EGP}
J. Balog, L. O'Raifeartaigh, P. Forgacs  and A. Wipf,
Nucl. Phys. {\bf B325} (1989) 225;
P. Petropoulos, 
Phys. Lett. {\bf B236} (1990) 151;
S. Hwang,
Nucl. Phys. {\bf B354} (1991) 100;
J. Evans, M. Gaverdiel and M. Perry,
Nucl. Phys. {\bf B535} (1998) 152,
hep-th/9806024.  

\bibitem{Matrix}
R. Dijkgraaf, E. Verlinde, H. Verlinde,
%{\it ``Matrix String Theory''},
Nucl. Phys. {\bf B500} (1997) 43,
hep-th/9703030;
L. Motl, 
%{\it ``Proposals on nonperturbative superstring
% interactions''},
hep-th/9701025;
T. Banks and N. Seiberg,
%{\it ``Strings form Matrices''}
hep-th/9702187.



\end{thebibliography}
\end{document}